\documentclass[11pt,a4paper]{article}

\usepackage[utf8]{inputenc}
\usepackage[T1]{fontenc}
\usepackage[english]{babel}
\usepackage{amsmath,amssymb,bm}
\usepackage{siunitx}
\usepackage{geometry}
\usepackage{booktabs}
\usepackage{longtable}
\usepackage{array}
\usepackage{xcolor}
\usepackage{hyperref}
\usepackage{enumitem}
\usepackage{comment}
\usepackage{graphicx}
\usepackage{subcaption}
\hypersetup{
  colorlinks=true,
  linkcolor=blue!55!black,
  citecolor=blue!55!black,
  urlcolor=blue!55!black
}
\setlist[itemize]{topsep=3pt,itemsep=2pt,parsep=2pt}
\setlist[enumerate]{topsep=3pt,itemsep=2pt,parsep=2pt}
\newcommand{\Tint}{T_{\mathrm{int}}}
\newcommand{\Tcom}{T_{\mathrm{COM}}}
\newcommand{\Tgas}{T_{\mathrm{gas}}}
\newcommand{\Trad}{T_{\mathrm{rad}}}
\newcommand{\Tbath}{T_{\mathrm{bath,COM}}}
\newcommand{\kb}{k_{\mathrm{B}}}

\newcommand{\Vtot}{V_{\mathrm{tot}}}
\newcommand{\Vact}{V_{\mathrm{act}}}
\newcommand{\Atot}{A_{\mathrm{tot}}}
\newcommand{\Aact}{A_{\mathrm{act}}}
\newcommand{\Cth}{C_{\mathrm{th}}}
\newcommand{\Pcool}{P_{\mathrm{cool,net}}}
\newcommand{\Pabsact}{P_{\mathrm{abs,active}}}
\newcommand{\Pabsbg}{P_{\mathrm{abs,bg}}}

\newcommand{\lambdaf}{\lambda_{\mathrm{f}}}
\newcommand{\etaext}{\eta_{\mathrm{ext}}}
\newcommand{\etac}{\eta_{\mathrm{c}}}
\newcommand{\GammaGas}{\Gamma_{\mathrm{gas}}}
\newcommand{\GammaInt}{\Gamma_{\mathrm{int}}}

\usepackage{authblk}

\title{Internal Temperature and Center-of-Mass Dynamics of a Levitated Nanoparticle: A Simulator}

\author[1,2]{Jailson Sales Araújo\thanks{ \texttt{jailson.araujo52.js@gmail.com}}}
\author[1,2]{Lucas Rodrigo Mendicino}
\author[1,2]{Lina Wölcken}
\author[1,2]{Santiago Gliosca}
\author[1,2]{Christian Tomás Schmiegelow}

\affil[1]{\textit{Universidad de Buenos Aires, Facultad de Ciencias Exactas y Naturales, Departamento de Física. Buenos Aires, Argentina}}
\affil[2]{\textit{ CONICET - Universidad de Buenos Aires, Instituto de Física de Buenos Aires (IFIBA). Buenos Aires, Argentina}}
\date{\today}

\begin{document}
\maketitle

\begin{abstract}
This document describes the models used in the simulator for the dynamics of a nanoparticle levitated in a Paul trap combined with an optical tweezer. The focus is the coupled evolution between the center-of-mass motion, the internal temperature of the particle, and the effective temperature associated with the mechanical motion. The formulation includes electric and optical forces, gas damping, thermal noise, heat exchange with the gas, thermal radiation, active optical absorption, parasitic absorption, anti-Stokes cooling, and absorption/emission spectra obtained from generic presets, generated bands, or experimental tables loaded by the user.
\end{abstract}

\tableofcontents
\newpage

\section{Introduction}

The physics of levitated nanoparticles naturally involves the interaction between mechanical dynamics, optical response, dissipation in a rarefied gas, and internal thermodynamics \cite{paul1990,ashkin1986,neuman2004,seletskiy2016,dani2021}. Even when the particle is treated as a classical object, the problem remains rich because different degrees of freedom respond to distinct physical mechanisms and to different time scales. The nanoparticle center of mass behaves as a charged oscillator subject to time-dependent electric fields; the optical response depends on the polarizability, size, and effective refractive index of the particle; the mechanical energy is dissipated and excited by collisions with gas molecules; and the internal energy can be increased or reduced by optical absorption, anti-Stokes fluorescence, thermal emission, and heat exchange with the environment. The interest of the system lies precisely in the fact that these sectors are not independent: the optical field that modifies the mechanical potential also deposits or removes internal energy; the internal temperature can alter the effective bath felt by the center of mass; and the gas pressure simultaneously controls mechanical dissipation and thermal exchange.

In a Paul trap, center-of-mass confinement does not arise from a static electrostatic minimum in all directions. Such a minimum would be incompatible with Laplace's equation in a region without free charges. Stability is obtained dynamically through an oscillating quadrupole field. At each instant, the field may be confining in some directions and deconfining in others, but the rapid temporal alternation produces an effective average confinement. The resulting dynamics is described by Mathieu-type equations and, within the stability regions, can be interpreted as the combination of a slow secular oscillation with a fast micromotion \cite{paul1990}.

When the nanoparticle is also illuminated by a focused laser beam, a second contribution to the mechanical potential appears. The optical tweezer produces gradient forces associated with the spatial variation of intensity and radiation-pressure forces associated with momentum transfer by absorption and scattering. In the regime in which the effective polarizability is positive, the gradient force tends to pull the particle toward regions of higher intensity, while the scattering force tends to push it along the propagation direction of the beam. Thus, the optical tweezer can increase the effective stiffness of the confinement, shift the equilibrium position, introduce asymmetries between axes, and modify the micromotion amplitude if the optical focus does not coincide with the center of the trap \cite{ashkin1986,ashkin1992,harada1996,neuman2004}. In the simulator, this optical contribution has another important role: the same local intensity that determines the optical force also enters the calculation of absorbed power, parasitic heating, and possible internal optical cooling.

The internal thermal description adds another layer to the problem. The internal temperature of the particle, denoted by $T_{\mathrm{int}}$, effectively represents the energy stored in the internal degrees of freedom of the material, such as lattice vibrational modes, thermal excitations, and electronic populations that thermalize with the solid. In materials with high external quantum efficiency and low parasitic absorption, optical pumping on the red side of the absorption band can produce anti-Stokes fluorescence: the particle absorbs lower-energy photons and emits, on average, higher-energy photons, extracting the energy difference from the lattice in the form of phonons \cite{pringsheim1929,landau1946,epstein1995,ruan2007,sheik2009,nemova2010,seletskiy2010,seletskiy2016}. This mechanism can reduce $T_{\mathrm{int}}$, but only if the power removed by fluorescence exceeds the competing heating channels. In practice, the internal temperature is determined by a balance between active absorption, parasitic absorption, fluorescent emission, infrared thermal radiation, and heat exchange with the residual gas.

It is important to emphasize that $T_{\mathrm{int}}$ should not be confused with the temperature associated with the center-of-mass motion, $T_{\mathrm{COM}}$. The former is an internal material temperature, connected to the thermal content of the nanoparticle. The latter is an effective or motional temperature, inferred from the translational kinetic energy or from the spectral width of the motion. These two temperatures may have very different values. A particle may be internally heated by optical absorption and, at the same time, have its center of mass cooled by damping or feedback. Conversely, the center of mass may gain mechanical energy through noise, micromotion, or trap instability while the internal temperature remains close to the ambient temperature. For this reason, the simulator treats $T_{\mathrm{int}}$ and $T_{\mathrm{COM}}$ as conceptually distinct variables, although it allows a phenomenological coupling between them.

The formulation presented in this document organizes the problem into three strongly connected sectors. The first is the mechanical dynamics of the center of mass, governed by the Paul trap, optical forces, compensation fields, Coulomb interactions when present, and dissipative or stochastic terms. The second is the internal thermal dynamics, described by an energy-balance equation that includes gas exchange, thermal radiation, optical absorption, fluorescent emission, and parasitic channels. The third is the coupling between the sectors, where the local optical intensity connects position and internal temperature, while the internal temperature can contribute to the effective mechanical bath felt by the center of mass.

This separation is not only a matter of presentation. It reflects the physical structure of the system and the approximations needed to make it tractable. The center of mass is described by classical position and velocity variables evolving on time scales associated with the secular frequency and the trap oscillator field. The internal temperature is described by an effective variable, assuming that the relevant internal degrees of freedom thermalize fast enough for a material temperature to be well defined. Exchange with the gas and radiation is treated by effective conductance or net-power models. The optical spectral response can be represented by phenomenological presets, generated bands, or experimental data loaded from a table. This approach makes it possible to study the competition between mechanisms without requiring a complete microscopic simulation of all internal modes, molecular collisions, and optical transitions.

Three temperatures repeatedly appear throughout the model. The gas temperature, $T_g$, characterizes the external molecular reservoir that produces mechanical damping and heat exchange through collisions. The internal temperature, $T_{\mathrm{int}}$, characterizes the effective thermal state of the nanoparticle. The center-of-mass temperature, $T_{\mathrm{COM}}$, characterizes the mechanical translational energy scale. At higher pressures, the gas tends to bring both the internal temperature and the motional temperature closer to $T_g$. At lower pressures, the connection with the gas weakens, and the dynamics becomes more sensitive to optical absorption, thermal emission, laser cooling, feedback, and residual noise. In regimes of efficient anti-Stokes cooling, $T_{\mathrm{int}}$ can decrease below ambient temperature while $T_{\mathrm{COM}}$ remains controlled by damping, noise, and trap stability. In regimes dominated by parasitic absorption, internal heating can in turn degrade mechanical stability if there is coupling between the internal degrees of freedom and the effective center-of-mass bath.

The physical scope of this document is therefore a levitated, charged, polarizable nanoparticle whose dynamics results from the competition between electrodynamic confinement, optical forces, dissipation in a rarefied gas, thermal noise, and internal energy balance. The goal is not merely to list terms used in a numerical implementation, but to build a physical description in which each contribution can be identified, interpreted, and evaluated within its validity regime. This perspective is essential for using the simulator critically: the implemented models are rich enough to explore physical trends, compare regimes, and test hypotheses, but their quantitative predictions for a specific material depend on the quality of the optical, thermal, and spectral parameters supplied by the user.

\subsection{Access to the interactive simulator}

This document was written to accompany the interactive simulator for nanoparticle dynamics in a Paul trap, optical tweezer, and laser cooling. The equations and models discussed in the following sections correspond to the physical mechanisms numerically implemented in the simulator. Thus, the reader can use the text as a theoretical reference and then access the simulator to modify parameters, visualize trajectories, monitor the evolution of $\Tint$ and $\Tcom$, analyze absorption and emission spectra, test different thermal regimes, and observe how each physical term alters the nanoparticle dynamics. The simulator, named \textit{Simulador Interativo de Nanopartículas Levitadas e Espectroscopia} (siNPle), is available online at
 \href{https://sinple-af59a.web.app}{\textbf{siNPle}}

\section{Mechanical dynamics of the center of mass}

The mechanical dynamics of the nanoparticle in the simulator describes the motion of its center of mass under the combined action of the Paul trap, the optical tweezer, possible compensation fields, and the dissipative/stochastic terms discussed in Section \ref{sec:dissipation_noise_and_temperature}. The fundamental mechanical variable is the position
\begin{equation}
  \bm{r}(t)=x(t)\hat{\bm{x}}+y(t)\hat{\bm{y}}+z(t)\hat{\bm{z}},
\end{equation}
and its associated velocity
\begin{equation}
  \bm{v}(t)=\frac{d\bm{r}}{dt}.
\end{equation}
The nanoparticle is treated as a classical, rigid object whose translational motion can be described by Newton's equations \cite{neuman2004,dani2021}. This means that the internal degrees of freedom of the particle, responsible for the internal temperature $\Tint$, are not mechanically resolved; they appear only through effective thermal couplings.

The general equation of motion can be written as
\begin{equation}
  m\frac{d^2\bm{r}}{dt^2}
  =
  \bm{F}_{\mathrm{Paul}}
  +
  \bm{F}_{\mathrm{opt}}
  +
  \bm{F}_{\mathrm{ext}}
  +
  \bm{F}_{\mathrm{drag}}
  +
  \bm{F}_{\mathrm{noise}},
  \label{eq:com_general_motion}
\end{equation}
where $m$ is the particle mass, $\bm{F}_{\mathrm{Paul}}$ is the electric force from the Paul trap, $\bm{F}_{\mathrm{opt}}$ is the optical force associated with the tweezer, $\bm{F}_{\mathrm{ext}}$ represents compensation fields or external displacements, $\bm{F}_{\mathrm{drag}}$ is the mechanical damping, and $\bm{F}_{\mathrm{noise}}$ is the stochastic force associated with the effective thermal bath.

\subsection{Mass, charge, and rigid-particle approximation}

For a homogeneous particle of radius $R$ and density $\rho$, the mass used in the center-of-mass dynamics is
\begin{equation}
  m = \frac{4\pi}{3}R^3\rho.
  \label{eq:mass_homogeneous_particle}
\end{equation}
The total charge of the particle is represented by
\begin{equation}
  Q = N_e e,
  \label{eq:particle_charge}
\end{equation}
where $N_e$ is the effective number of elementary charges and $e$ is the elementary charge. The sign of $N_e$ defines whether the particle responds as positively or negatively charged. The charge is the parameter that connects the particle to the electric field of the Paul trap. Therefore, changing $Q$ directly modifies the electric force and the mechanical frequencies associated with electrodynamic confinement.

\subsection{Paul trap: time-dependent electric field}

The Paul trap confines the charged particle by means of a time-dependent quadrupole electric potential \cite{paul1990}. The central physical idea is that a purely electrostatic field cannot produce stable confinement in all directions simultaneously, as a consequence of Earnshaw's theorem. The Paul trap circumvents this limitation by using an alternating oscillator field. At each instant, the potential may be defocusing in one direction and focusing in another, but the rapid temporal alternation produces an effective average confinement.

In a generic form, the electric potential near the trap center can be written as a quadratic expansion:
\begin{equation}
  \Phi_{\mathrm{Paul}}(\bm{r},t)
  =
  \frac{1}{2}
  \left[
  K_x(t)x^2+
  K_y(t)y^2+
  K_z(t)z^2
  \right],
  \label{eq:paul_quadratic_potential}
\end{equation}
with
\begin{equation}
  K_i(t)=K_i^{\mathrm{DC}}+K_i^{\mathrm{AC}}\cos(\Omega_{\mathrm{AC}}t),
  \qquad i=x,y,z.
  \label{eq:paul_curvatures}
\end{equation}
Here, $\Omega_{\mathrm{AC}}$ is the angular frequency of the oscillator field, where AC is alternating current. The coefficients $K_i^{\mathrm{DC}}$ represent static curvatures, while $K_i^{\mathrm{AC}}$ represent oscillating curvatures. Since the electrostatic potential must satisfy Laplace's equation in the charge-free region
\begin{equation}
  \nabla^2\Phi_{\mathrm{Paul}}=0,
\end{equation}
the curvatures are not independent; ideally
\begin{equation}
  K_x(t)+K_y(t)+K_z(t)=0.\label{eq:laplace_con}
\end{equation}
This expresses the fact that the field cannot be harmonically confining in all directions at the same instant.

The electric force on the charged particle is
\begin{equation}
  \bm{F}_{\mathrm{Paul}}
  =
  Q\bm{E}
  =
  -Q\nabla\Phi_{\mathrm{Paul}}.
\end{equation}
Using the quadratic potential in Eq.~\eqref{eq:paul_quadratic_potential}, we obtain
\begin{equation}
  F_{\mathrm{Paul},i}
  =
  -QK_i(t)r_i,
  \qquad i=x,y,z.
  \label{eq:paul_force_components}
\end{equation}
Therefore, in each axis, the force is proportional to the displacement relative to the trap center, but the effective spring constant oscillates in time.

The coefficients $K_i^{\mathrm{DC}}$ describe the static curvature generated mainly by the endcap electrodes, while $K_i^{\mathrm{AC}}$ describe the oscillating curvature produced by the AC electrodes. These coefficients have units of $\mathrm{V\,m^{-2}}$ and should not be confused with mechanical spring constants.  

In the quadrupole approximation, the AC contribution to the potential is given by
\begin{equation}
\Phi_{\mathrm{AC}}(x,y,t)
=
\frac{V_{\mathrm{AC}}}{2 r_0^2}
(x^2 - y^2)
\cos(\Omega_{\mathrm{AC}} t),
\end{equation}
where $V_{\mathrm{AC}}$ is the AC voltage amplitude and $r_0$ is the characteristic radial distance from the trap center to the AC electrodes. A purely time-dependent term has been omitted since it does not contribute to the electric field.

The static contribution from the endcap electrodes is
\begin{equation}
\Phi_{\mathrm{DC}}(x,y,z)
=
\frac{\kappa U_{\mathrm{ec}}}{z_0^2}
\left[
z^2 - \frac{x^2 + y^2}{2}
\right],
\end{equation}
where $U_{\mathrm{ec}}$ is the symmetric endcap voltage, $z_0$ is the characteristic axial length scale, and $\kappa$ is a dimensionless geometrical efficiency factor accounting for deviations from an ideal hyperbolic geometry.

The total potential is then
\begin{equation}
\Phi(\mathbf{r},t)
=
\frac{V_{\mathrm{AC}}}{2 r_0^2}
(x^2 - y^2)
\cos(\Omega_{\mathrm{AC}} t)
+
\frac{\kappa U_{\mathrm{ec}}}{z_0^2}
\left[
z^2 - \frac{x^2 + y^2}{2}
\right].
\end{equation}

From this expression, the time-dependent curvature coefficients can be identified. The DC curvatures are
\begin{equation}
K_x^{\mathrm{DC}} = K_y^{\mathrm{DC}} = -\frac{\kappa U_{\mathrm{ec}}}{z_0^2},
\qquad
K_z^{\mathrm{DC}} = \frac{2\kappa U_{\mathrm{ec}}}{z_0^2},
\end{equation}
while the AC curvatures are
\begin{equation}
K_x^{\mathrm{AC}} = \frac{V_{\mathrm{AC}}}{r_0^2},
\qquad
K_y^{\mathrm{AC}} = -\frac{V_{\mathrm{AC}}}{r_0^2},
\qquad
K_z^{\mathrm{AC}} = 0.
\end{equation}

Thus, the full time-dependent coefficients are
\begin{align}
K_x(t) &= -\frac{\kappa U_{\mathrm{ec}}}{z_0^2}
+ \frac{V_{\mathrm{AC}}}{r_0^2} \cos(\Omega_{\mathrm{AC}} t), \\
K_y(t) &= -\frac{\kappa U_{\mathrm{ec}}}{z_0^2}
- \frac{V_{\mathrm{AC}}}{r_0^2} \cos(\Omega_{\mathrm{AC}} t), \\
K_z(t) &= \frac{2\kappa U_{\mathrm{ec}}}{z_0^2}.
\end{align}

By using Eq.\eqref{eq:laplace_con}, we can write separately for DC and AC contributions,
\begin{equation}
K_x^{\mathrm{DC}} + K_y^{\mathrm{DC}} + K_z^{\mathrm{DC}} = 0,
\qquad
K_x^{\mathrm{AC}} + K_y^{\mathrm{AC}} + K_z^{\mathrm{AC}} = 0.
\end{equation}

For asymmetric endcap voltages, the combination
\begin{equation}
\frac{U_{\mathrm{up}} - U_{\mathrm{down}}}{2}
\end{equation}
introduces a linear axial field that shifts the equilibrium position. After expansion around the displaced equilibrium, the quadratic form remains valid.

More generally, the curvature coefficients can be defined from the Hessian of the potential,
\begin{equation}
K_i(t)
=
\left.
\frac{\partial^2 \Phi(\mathbf{r},t)}{\partial x_i^2}
\right|_{\mathbf{r} = \mathbf{r}_{\mathrm{eq}}}.
\end{equation}

Finally, $V_{\mathrm{AC}}$ denotes a peak voltage amplitude. If a peak-to-peak voltage $V_{\mathrm{pp}}$ is used instead, the conversion is
\begin{equation}
V_{\mathrm{AC}} = \frac{V_{\mathrm{pp}}}{2}.
\end{equation}
This distinction is important because the effective pseudopotential scales as $V_{\mathrm{AC}}^2$.

\subsection{Mathieu equations and secular stability}

Temporarily ignoring optical forces, damping, and noise, the equation of motion in each axis becomes
\begin{equation}
  \frac{d^2 r_i}{dt^2}
  +
  \frac{Q}{m}
  \left[
  K_i^{\mathrm{DC}}+
  K_i^{\mathrm{AC}}\cos(\Omega_{\mathrm{AC}}t)
  \right]r_i
  =
  0.
  \label{eq:paul_axis_motion}
\end{equation}
This is a physical form of the Mathieu equation \cite{paul1990}. To make this connection explicit, one introduces the dimensionless variable
\begin{equation}
  \tau=\frac{\Omega_{\mathrm{AC}}t}{2}.
\end{equation}
Eq.~\eqref{eq:paul_axis_motion} can be written as
\begin{equation}
  \frac{d^2 r_i}{d\tau^2}
  +
  \left[
  a_i
  +
  2q_i\cos(2\tau)
  \right]r_i
  =
  0,
  \label{eq:mathieu_form}
\end{equation}
where the dimensionless parameters are, equivalently,
\begin{align}
  a_i &=
  \frac{4QK_i^{\mathrm{DC}}}{m\Omega_{\mathrm{AC}}^2},
  \\
  q_i &=
  \frac{2QK_i^{\mathrm{AC}}}{m\Omega_{\mathrm{AC}}^2}.
\end{align}
The parameters $a_i$ and $q_i$ determine whether the motion is stable or unstable. When the solution is stable, the motion can be interpreted as the superposition of two time scales: a slow oscillation, called secular motion, and a fast oscillation at the oscillator field, called micromotion \cite{paul1990,dani2021}.

In the usual stability regime, for $|a_i|\ll1$ and $q_i^2\ll1$, the approximate secular frequency is
\begin{equation}
  \omega_i
  \simeq
  \frac{\Omega_{\mathrm{AC}}}{2}
  \sqrt{
  a_i+\frac{q_i^2}{2}
  }.
  \label{eq:secular_frequency}
\end{equation}
This expression shows that even when there is no strong static confinement, the AC term can produce an effective secular frequency proportional to $q_i\Omega_{\mathrm{AC}}$. In the simulator, this approximation helps interpret stability maps and typical oscillation scales, but the time dynamics can be integrated directly using the time-dependent field.


\subsection{Effective pseudopotential}

In many regimes, the oscillator field is much faster than the secular motion \cite{paul1990}. In this case, it is useful to describe the average effect of the AC by an effective pseudopotential. The average potential energy associated with secular confinement can be written as
\begin{equation}
  U_{\mathrm{pseudo}}(\bm{r})
  =
  \frac{1}{2}m
  \left(
  \omega_x^2x^2+
  \omega_y^2y^2+
  \omega_z^2z^2
  \right),
  \label{eq:pseudopotential}
\end{equation}
where $\omega_i$ are the secular frequencies. The corresponding average force is
\begin{equation}
  \bm{F}_{\mathrm{pseudo}}
  =
  -\nabla U_{\mathrm{pseudo}}
  =
  -m
  \left(
  \omega_x^2x\hat{\bm{x}}+
  \omega_y^2y\hat{\bm{y}}+
  \omega_z^2z\hat{\bm{z}}
  \right).
\end{equation}

This description is useful for understanding the global stability of the trap. However, it does not explicitly contain micromotion. When the particle is displaced from the AC null, micromotion can become significant. This point is important because compensation fields, optical forces, or misalignments can shift the mean position of the particle and increase the micromotion amplitude.

\subsection{Compensation fields and equilibrium displacement}

The simulator allows the inclusion of fields or displacements that move the equilibrium position of the particle \cite{paul1990,dani2021}. An approximately uniform electric field,
\begin{equation}
  \bm{E}_{\mathrm{comp}}
  =
  E_x\hat{\bm{x}}+
  E_y\hat{\bm{y}}+
  E_z\hat{\bm{z}},
\end{equation}
produces a force
\begin{equation}
  \bm{F}_{\mathrm{ext}}
  =
  Q\bm{E}_{\mathrm{comp}}.
\end{equation}
This term does not change the trap curvature, but it shifts the equilibrium point. In a harmonically approximated trap, the mean displacement in each axis can be estimated as
\begin{equation}
  r_{0,i}
  \simeq
  \frac{Q E_i}{m\omega_i^2}.
\end{equation}
This expression shows that weak fields can produce relevant displacements when the secular frequency is small or when the particle charge is high.

There may also be displacements associated with the optical tweezer itself. If the tweezer focus does not coincide with the electric center of the trap, the optical force pulls the particle toward the focus, while the Paul trap tries to keep it near the electrodynamic center. The final equilibrium results from the competition between these two potentials.

\subsection{Optical tweezer: local intensity and gradient force}

The optical tweezer acts on the nanoparticle through the interaction between the electromagnetic field of the laser and the effective polarizability of the material \cite{ashkin1986,ashkin1992,harada1996,neuman2004}. In the regime where the particle is sufficiently small compared with the wavelength in the medium, the optical response can be approximated by that of an induced dipole. In this limit, the electric field of the laser polarizes the particle, and the particle experiences a force associated with the spatial variation of the optical intensity. For a dielectric particle with positive polarizability relative to the external medium, this force points toward regions of higher intensity. This is the physical origin of the gradient force and of transverse optical confinement in a Gaussian tweezer.

In the simulator, each optical tweezer is treated in a local coordinate system associated with the beam itself. The coordinate $s$ measures the position along the propagation direction, while $v$ and $w$ are transverse coordinates. The transverse distance to the optical axis is
\begin{equation}
  \rho^2=v^2+w^2.
\end{equation}
For a Gaussian beam of power $P$, the local intensity is written as
\begin{equation}
  I(\rho,s)
  =
  \frac{2P}{\pi w^2(s)}
  \exp\left[
  -\frac{2\rho^2}{w^2(s)}
  \right],
  \label{eq:gaussian_intensity}
\end{equation}
where $w(s)$ is the beam radius at axial position $s$:
\begin{equation}
  w(s)=w_0
  \sqrt{
  1+\left(\frac{s}{z_R}\right)^2
  }.
\end{equation}
The Rayleigh length is
\begin{equation}
  z_R=\frac{\pi n_m w_0^2}{\lambda_L},
\end{equation}
where $n_m$ is the refractive index of the external medium and $\lambda_L$ is the laser wavelength in vacuum. The waist $w_0$ determines the spatial scale of the focus: the smaller $w_0$ is, the higher the focal intensity and the stronger the optical gradient near the center of the beam.

In the Rayleigh regime, the effective polarizability of a homogeneous sphere can be written as \cite{bohren1983,harada1996}
\begin{equation}
  \alpha_p
  =
  4\pi\epsilon_0 n_m^2 R^3
  \left(
  \frac{m_r^2-1}{m_r^2+2}
  \right),
  \qquad
  m_r=\frac{n_p}{n_m},
  \label{eq:rayleigh_polarizability}
\end{equation}
where $R$ is the particle radius, $n_p$ is the refractive index of the particle, and $m_r$ is the relative refractive index. This expression shows that, in the dipole limit, the optical force amplitude grows with the particle volume, since $\alpha_p\propto R^3$. It also shows that the sign of the gradient force depends on the optical contrast between the particle and the medium. When $n_p>n_m$, usually $\alpha_p>0$, and the particle is attracted to the intensity maximum. If the effective polarizability were negative, the gradient force would have the opposite direction and the particle would tend to be repelled from the most intense regions.

The dipole potential energy associated with the optical field can be written, in an effective form, as
\begin{equation}
  U_{\mathrm{opt}}(\bm{r})
  =
  -\frac{\alpha_p}{2n_m\epsilon_0 c}
  I(\bm{r}),
  \label{eq:optical_potential}
\end{equation}
and the gradient force follows from
\begin{equation}
  \bm{F}_{\mathrm{grad}}
  =
  -\nabla U_{\mathrm{opt}}
  =
  \frac{\alpha_p}{2n_m\epsilon_0 c}
  \nabla I(\bm{r}).
  \label{eq:gradient_force}
\end{equation}
Thus, the gradient force is determined by two quantities: the polarizability of the particle and the spatial gradient of the intensity. In the transverse plane, the Gaussian profile provides a restoring force toward the beam axis. In the axial direction, the gradient force arises because the beam radius and the maximum intensity vary along $s$. This axial contribution is weaker than the transverse contribution in many regimes, but it is essential for the existence of three-dimensional confinement in a single-beam optical tweezer.

In the local beam frame, the conservative part of the optical force can be represented by
\begin{align}
  F_v^{\mathrm{grad}} &= c_{\mathrm{dip}}\frac{\partial I}{\partial v},\\
  F_w^{\mathrm{grad}} &= c_{\mathrm{dip}}\frac{\partial I}{\partial w},\\
  F_s^{\mathrm{grad}} &= c_{\mathrm{dip}}\frac{\partial I}{\partial s},
\end{align}
with
\begin{equation}
  c_{\mathrm{dip}}
  =
  \frac{\alpha_p}{2n_m\epsilon_0 c}.
  \label{eq:cdip}
\end{equation}
These expressions make clear that the gradient force depends not only on the total laser power, but on the local intensity experienced by the particle. Therefore, when the particle is displaced within the Gaussian profile, the force changes automatically because both $I$ and its spatial derivatives change.

\subsection{Scattering force and radiation pressure}

The description of the optical tweezer is not complete with the gradient force alone. Light also carries linear momentum and can transfer part of this momentum to the particle through scattering and absorption \cite{ashkin1992,harada1996,bohren1983}. This transfer produces a force component oriented, on average, along the beam propagation direction. This component is called the scattering force, radiation-pressure force, or non-conservative force of the optical tweezer.

The physical origin of radiation pressure lies in momentum conservation. A beam of intensity $I$ carries a momentum flux proportional to $n_m I/c$. If the particle removes power from the original incident mode, the electromagnetic field loses momentum in the propagation direction. This loss appears as a mechanical force on the particle. In a general formulation, the force associated with extinction of the beam can be written as
\begin{equation}
  \bm{F}_{\mathrm{scatt}}
  \simeq
  \frac{n_m}{c}
  C_{\mathrm{ext}} I(\bm{r})\hat{\bm{k}},
  \label{eq:scattering_force}
\end{equation}
where $\hat{\bm{k}}$ is the unit vector in the propagation direction and $C_{\mathrm{ext}}$ is an effective extinction cross section. Extinction represents the power removed from the incident beam and, in a complete electromagnetic description, includes both scattering and absorption,
\begin{equation}
  C_{\mathrm{ext}}
  \simeq
  C_{\mathrm{scatt}}+C_{\mathrm{abs}}.
\end{equation}
This distinction is important because scattering and absorption have different physical consequences. Scattering redirects light and transfers momentum to the particle without necessarily converting optical energy into internal heat. Absorption also transfers momentum, but it deposits energy in the particle and therefore contributes directly to the thermal balance of $\Tint$.

In the Rayleigh limit, the scattering cross section of a dielectric sphere can be approximated by
\begin{equation}
  \sigma_{\mathrm{scatt}}
  =
  \frac{8\pi}{3}
  k_m^4R^6
  \left(
  \frac{m_r^2-1}{m_r^2+2}
  \right)^2,
  \qquad
  k_m=\frac{2\pi n_m}{\lambda_L}.
  \label{eq:rayleigh_scattering_cross_section}
\end{equation}
Comparing this expression with the polarizability in Eq.~\eqref{eq:rayleigh_polarizability}, one sees a fundamental scaling difference: the gradient force is proportional to the polarizability and therefore grows as $R^3$, while the radiation-pressure force associated with scattering grows as $\sigma_{\mathrm{scatt}}\propto R^6$. For this reason, for very small and weakly absorbing particles, the gradient force usually dominates. For larger, more refractive, more absorbing particles, or under high intensities, the scattering force can become comparable to the gradient force and strongly modify the stability of the tweezer.

In the simulator, the total mechanical contribution of the optical tweezer is written in the local beam frame as the sum of the gradient force and the scattering radiation pressure:
\begin{align}
  F_v &= c_{\mathrm{dip}}\frac{\partial I}{\partial v},\\
  F_w &= c_{\mathrm{dip}}\frac{\partial I}{\partial w},\\
  F_s &= c_{\mathrm{dip}}\frac{\partial I}{\partial s}
  +
  \frac{n_m}{c}\sigma_{\mathrm{scatt}} I.
  \label{eq:local_optical_force_components}
\end{align}
The components $F_v$ and $F_w$ are purely transverse and restoring when $\alpha_p>0$, because they pull the particle toward the beam axis. The component $F_s$ contains two terms of different nature. The first is the axial gradient force, which depends on the longitudinal variation of the intensity. The second is radiation pressure, proportional to the local intensity and oriented in the direction of propagation. Consequently, even when the axial gradient force changes sign when crossing the focus, radiation pressure continues to push the particle forward.

This asymmetry explains why the axial confinement of a single-beam optical tweezer is more delicate than the transverse confinement. Transversely, the particle is pulled toward the center of the Gaussian profile. Axially, the gradient force must compensate radiation pressure. When radiation pressure is small, the equilibrium position lies close to the focus, usually shifted in the propagation direction. When radiation pressure increases, this equilibrium point moves away from the focus and the effective axial stiffness decreases. If the scattering term becomes larger than the maximum restoring axial force, the tweezer no longer confines the particle along the optical axis.

The difference between Eq.~\eqref{eq:scattering_force} and Eq.~\eqref{eq:local_optical_force_components} also clarifies the approximation used in the simulator. The general expression uses $C_{\mathrm{ext}}$ because, in principle, both absorption and scattering remove momentum from the incident beam. The mechanical implementation of the tweezer, however, explicitly uses the Rayleigh scattering cross section, $\sigma_{\mathrm{scatt}}$, for radiation pressure. Active optical absorption and parasitic absorption are not treated as an additional mechanical force in this part of the model; they enter mainly in the internal-temperature balance through the absorbed power and the optical heating or cooling terms.

This separation between mechanical action and thermal action is essential for interpreting the simulator. The same local intensity $I(\bm{r})$ that determines the gradient force and radiation pressure also determines the optical power absorbed by the particle. Therefore, when the particle moves within the Gaussian profile, both its mechanical acceleration and its internal heating or cooling rate change simultaneously. If the material is in an efficient anti-Stokes cooling regime, a higher local intensity can increase the cooling of $\Tint$. If parasitic absorption or nonradiative losses dominate, the same higher intensity increases heating. Thus, the optical tweezer in the simulator should be understood as a coupled optomechanical and thermo-optical channel, in which position, force, local intensity, and internal thermal balance are not independent.

\subsection{Harmonic approximation near the optical focus}

Near the focus, the beam intensity can be expanded around the equilibrium position \cite{ashkin1986,neuman2004}. For small displacements, the optical force can be approximated by a restoring force
\begin{equation}
  F_{\mathrm{opt},i}
  \simeq
  -k_i^{\mathrm{opt}}
  \left(r_i-r_{f,i}\right),
  \label{eq:optical_harmonic_force}
\end{equation}
where $\bm{r}_f$ is the focus position and $k_i^{\mathrm{opt}}$ is the effective optical stiffness along axis $i$. This stiffness depends on the laser power, the waist, the particle polarizability, and the beam geometry.

The mechanical frequency associated with the isolated optical confinement would be
\begin{equation}
  \omega_i^{\mathrm{opt}}
  =
  \sqrt{\frac{k_i^{\mathrm{opt}}}{m}}.
\end{equation}
When the optical tweezer and the Paul trap are active simultaneously, the restoring contributions add approximately in the linear regime \cite{paul1990,ashkin1986,dani2021}
\begin{equation}
  \omega_{i,\mathrm{eff}}^2
  \simeq
  \omega_{i,\mathrm{Paul}}^2
  +
  \omega_{i,\mathrm{opt}}^2,
  \label{eq:combined_frequency}
\end{equation}
provided that the centers of the potentials are close and nonlinearities are small. If the centers are displaced, the effective frequency may still increase, but the equilibrium point changes and the particle may experience additional micromotion or explore non-harmonic regions of the potential.

\subsection{Combination of Paul trap and optical tweezer}

The total center-of-mass dynamics results from the sum of electric and optical forces \cite{paul1990,ashkin1986,dani2021}, could be write as 
\begin{equation}
  m\frac{d^2\bm{r}}{dt^2}
  =
  \bm{F}_{\mathrm{Paul}}(\bm{r},t)
  +
  \bm{F}_{\mathrm{opt}}(\bm{r})
  +
  \bm{F}_{\mathrm{ext}}
  + \cdots,
  \label{eq:combined_mechanical_dynamics}
\end{equation}
where the dots denote dissipative and stochastic terms. This sum can produce different physical regimes.

If the Paul trap dominates, the optical tweezer acts mainly as a local perturbation, slightly shifting the equilibrium and modifying the secular frequencies. If the optical tweezer dominates, the motion can become strongly localized near the optical focus, while the electric trap provides additional long-range confinement. In intermediate regimes, the competition between AC confinement, optical gradient, scattering, and damping can produce complex trajectories.

This hybrid structure is useful because it separates two physical functions, the Paul trap provides a deep, large-volume potential, while the optical tweezer provides local control, a strong intensity gradient, and direct coupling to optical absorption, emission, and heating processes \cite{paul1990,neuman2004,dani2021}.

\subsection{Micromotion and secular motion}

In the Paul trap, even when the motion is stable, the trajectory is not simply a pure harmonic oscillation \cite{paul1990}. It contains a slow component and a fast component. The slow component is the secular motion, associated with the frequencies $\omega_i$. The fast component is the micromotion, directly forced by the AC field.

Approximately, in a radial axis, the solution can be written as
\begin{equation}
  r_i(t)
  \simeq
  R_i(t)
  \left[
  1+\frac{q_i}{2}\cos(\Omega_{\mathrm{AC}}t)
  \right],
  \label{eq:micromotion_approx}
\end{equation}
where $R_i(t)$ varies slowly on the secular time scale. This expression shows that the micromotion amplitude increases with the Mathieu parameter $q_i$ and with the secular displacement relative to the AC null.

When external forces displace the particle away from the AC center, excess micromotion appears \cite{paul1990,dani2021}. This is relevant for the optical tweezer because a misaligned focus can pull the particle to a position where the AC field does not vanish. In this situation, even if the secular motion appears well confined, the particle can exhibit an additional fast oscillation, increasing the instantaneous kinetic energy and affecting the readout of $\Tcom$.

\subsection{Effective mechanical energy}

To interpret mechanical stability, it is useful to define an effective mechanical energy \cite{paul1990,hairer2003}
\begin{equation}
  E_{\mathrm{mech}}
  =
  \frac{1}{2}m|\bm{v}|^2
  +
  U_{\mathrm{pseudo}}(\bm{r})
  +
  U_{\mathrm{opt}}(\bm{r}).
  \label{eq:effective_mechanical_energy}
\end{equation}
This expression is approximate because the real potential of the Paul trap is time dependent. Therefore, the mechanical energy is not exactly conserved even without damping. Even so, $E_{\mathrm{mech}}$ is useful for understanding whether the particle is localized in a stable region of the potential or whether it is exploring regions from which it may escape.

The kinetic part,
\begin{equation}
  E_{\mathrm{kin}}=\frac{1}{2}m|\bm{v}|^2,
\end{equation}
is the basis for the definition of the center-of-mass kinetic temperature discussed in Section \ref{sec:dissipation_noise_and_temperature}. The potential terms indicate the effective trap depth. When the mechanical energy approaches the confinement depth, the particle can escape, especially if there is heating by noise, excessive micromotion, or non-restoring optical forces.

\section{Dissipation, thermal noise, and center-of-mass temperature}
\label{sec:dissipation_noise_and_temperature}
The center-of-mass dynamics of the nanoparticle is not determined only by the conservative forces of the Paul trap and the optical tweezer \cite{paul1990,neuman2004,dani2021}. In a real gaseous environment, even at low pressure, the particle undergoes collisions with residual gas molecules. These collisions produce two complementary effects: an average dissipative force, which removes mechanical energy from the motion, and a fluctuating force, which injects thermal noise into the center of mass.

In the simulator, these effects are treated by an effective Langevin-type description \cite{kloeden1992,higham2001}. Dissipation appears as a damping term proportional to velocity, while thermal noise is introduced as random velocity increments compatible with the effective mechanical bath. This approach does not resolve individual molecular collisions, but it captures the average statistical effect of the gas on the translational kinetic energy of the nanoparticle.

The center-of-mass temperature, $\Tcom$, must be interpreted carefully. It is not the internal temperature of the nanoparticle. The internal temperature $\Tint$ describes the energy stored in the vibrational and thermal degrees of freedom of the material. By contrast, $\Tcom$ is an effective measure of the kinetic energy associated with the translational motion of the particle in the trap. Thus, a particle may have very high $\Tint$ and low $\Tcom$, or vice versa, depending on the heating, cooling, and coupling channels present.

\subsection{Gas damping of the center of mass}

The main mechanical dissipative channel considered in the simulator is damping by residual gas. For small particles in the molecular or intermediate regime, a common approximation is the Epstein form, in which the damping rate depends on the gas density, the mean thermal velocity of the molecules, and the geometric properties of the particle \cite{epstein1924,hebestreit2018}. The simulator uses the effective form
\begin{equation}
  \GammaGas = \frac{\rho_g\bar{v}_g C_E}{\rho R},
\end{equation}
where $\rho$ is the material density of the nanoparticle, $R$ is the external radius, and $C_E$ is a numerical Epstein factor. The gas quantities are calculated as
\begin{align}
  \rho_g &= \frac{P m_g}{\kb \Tgas}, \\
  \bar{v}_g &= \sqrt{\frac{8\kb\Tgas}{\pi m_g}}, \\
  C_E &= 1+\frac{\pi}{8}.
\end{align}

Here $P$ is the pressure, $m_g$ is the effective molecular mass of the gas, $\Tgas$ is the gas temperature, and $\bar{v}_g$ is the mean thermal velocity of the molecules. The pressure dependence appears through $\rho_g$. Therefore, when the pressure is reduced, the molecular density decreases and gas damping becomes smaller. This means that, in high vacuum, the particle preserves its mechanical energy for a longer time, but it also becomes less coupled to the thermal bath of the gas.

The rate $\GammaGas$ enters the equation of motion as a dissipative force, given by
\begin{equation}
  \bm{F}_{\mathrm{drag}}=-m\GammaGas\bm{v},
\end{equation}
where $m$ is the nanoparticle mass and $\bm{v}$ is the center-of-mass velocity. This term always opposes the instantaneous motion and therefore removes translational kinetic energy. In a purely deterministic description, this damping would bring the particle to rest at the effective potential minimum. Physically, however, the same gas that dissipates also produces thermal fluctuations, so the motion does not relax to zero energy, but to an energy compatible with the bath temperature.

\subsection{Thermal noise and Langevin interpretation}
\label{subsec:thermal_noise}

The simulator describes the stochastic center-of-mass motion of the nanoparticle using a Langevin model. Two effective mechanical reservoirs are considered. The first is the surrounding gas, which produces viscous damping and random momentum transfer through molecular collisions. The second is an optional phenomenological reservoir that couples the nanoparticle's internal temperature \(T_{\mathrm{int}}\) to its center-of-mass motion. These two contributions must be distinguished from the optical and feedback forces, which are treated deterministically in the current implementation.

For a nanoparticle of mass \(m\), the translational dynamics can be written as
\begin{equation}
    m\frac{d\bm{v}}{dt}
    =
    \bm{F}_{\mathrm{Paul}}
    +
    \bm{F}_{\mathrm{opt}}
    +
    \bm{F}_{\mathrm{C}}
    +
    \bm{F}_{\mathrm{ext}}
    +
    \bm{F}_{\mathrm{fb}}
    -
    m\Gamma_{\mathrm{phy}}\bm{v}
    +
    \bm{\xi}(t),
    \label{eq:langevin_com}
\end{equation}
where \(\bm{F}_{\mathrm{Paul}}\) is the force generated by the Paul trap, \(\bm{F}_{\mathrm{opt}}\) contains the optical gradient and scattering forces, \(\bm{F}_{\mathrm{C}}\) accounts for Coulomb interactions between charged nanoparticles, and \(\bm{F}_{\mathrm{ext}}\) represents any additional deterministic force. The feedback force is denoted by \(\bm{F}_{\mathrm{fb}}\), while \(\bm{\xi}(t)\) is the total stochastic mechanical force.

The total physical damping rate is
\begin{equation}
    \Gamma_{\mathrm{phy}}
    =
    \Gamma_{\mathrm{gas}}
    +
    \Gamma_{\mathrm{int}},
    \label{eq:total_physical_damping}
\end{equation}
where \(\Gamma_{\mathrm{gas}}\) is the damping rate produced by collisions with the background gas and \(\Gamma_{\mathrm{int}}\) is the phenomenological coupling rate between the internal and center-of-mass degrees of freedom. In the parametrization adopted by the simulator,
\begin{equation}
    \Gamma_{\mathrm{int}}
    =
    \chi_{\mathrm{int}\rightarrow\mathrm{COM}}
    \Gamma_{\mathrm{gas}},
    \label{eq:internal_com_coupling}
\end{equation}
where \(\chi_{\mathrm{int}\rightarrow\mathrm{COM}}\) controls the strength of this coupling. When the internal--COM coupling is disabled, \(\chi_{\mathrm{int}\rightarrow\mathrm{COM}}=0\) and consequently \(\Gamma_{\mathrm{int}}=0\).

The total stochastic force can be interpreted as the sum of two contributions,
\begin{equation}
    \bm{\xi}(t)
    =
    \bm{\xi}_{\mathrm{gas}}(t)
    +
    \bm{\xi}_{\mathrm{int}}(t).
    \label{eq:total_random_force}
\end{equation}
The gas contribution \(\bm{\xi}_{\mathrm{gas}}(t)\) represents the accumulated momentum transferred by many gas--particle collisions. The simulator does not resolve each molecular impact individually. Instead, the detailed sequence of collisions is replaced by Gaussian white noise, an approximation appropriate when the molecular correlation time is much shorter than the characteristic timescale of the nanoparticle motion.

The contribution \(\bm{\xi}_{\mathrm{int}}(t)\) represents the mechanical fluctuations associated with the phenomenological internal--COM coupling. In this description, the internal degrees of freedom behave as an effective mechanical reservoir at the instantaneous internal temperature \(T_{\mathrm{int}}\). It provides a simplified way of allowing changes in \(T_{\mathrm{int}}\) to influence the mechanical fluctuations and, consequently, the effective center-of-mass temperature.

For a reservoir \(\alpha\), with \(\alpha\in\{\mathrm{gas},\mathrm{int}\}\), the corresponding random force has zero mean,
\begin{equation}
    \left\langle
        \xi_{\alpha,i}(t)
    \right\rangle
    =0,
\end{equation}
and satisfies the fluctuation--dissipation relation
\begin{equation}
    \left\langle
        \xi_{\alpha,i}(t)
        \xi_{\beta,j}(t')
    \right\rangle
    =
    2m\kb\Gamma_{\alpha}T_{\alpha}\,
    \delta_{\alpha\beta}
    \delta_{ij}
    \delta(t-t'),
    \label{eq:reservoir_noise_correlation}
\end{equation}
where
\begin{equation}
    T_{\alpha}
    =
    \begin{cases}
        T_{\mathrm{gas}}, & \alpha=\mathrm{gas},\\
        T_{\mathrm{int}}, & \alpha=\mathrm{int}.
    \end{cases}
\end{equation}
Here \(i,j\in\{x,y,z\}\). The factor \(\delta_{ij}\) indicates that the random impulses applied along different Cartesian directions are statistically independent, while \(\delta_{\alpha\beta}\) indicates that the gas and internal reservoirs are uncorrelated. The Dirac delta \(\delta(t-t')\) expresses the white-noise approximation, according to which stochastic forces evaluated at different times are uncorrelated.

Combining the two reservoirs gives
\begin{equation}
    \left\langle
        \xi_i(t)\xi_j(t')
    \right\rangle
    =
    2m\kb
    \left(
        \Gamma_{\mathrm{gas}}T_{\mathrm{gas}}
        +
        \Gamma_{\mathrm{int}}T_{\mathrm{int}}
    \right)
    \delta_{ij}\delta(t-t').
    \label{eq:total_noise_correlation}
\end{equation}
Therefore, both reservoirs contribute to the mechanical diffusion. The gas contribution is determined by \(\Gamma_{\mathrm{gas}}T_{\mathrm{gas}}\), whereas the phenomenological internal contribution is determined by \(\Gamma_{\mathrm{int}}T_{\mathrm{int}}\).

It is convenient to define an effective center-of-mass bath temperature,
\begin{equation}
    T_{\mathrm{bath}}^{\mathrm{COM}}
    =
    \frac{
        \Gamma_{\mathrm{gas}}T_{\mathrm{gas}}
        +
        \Gamma_{\mathrm{int}}T_{\mathrm{int}}
    }{
        \Gamma_{\mathrm{gas}}
        +
        \Gamma_{\mathrm{int}}
    }.
    \label{eq:effective_com_bath_temperature}
\end{equation}
With this definition, Eq.~\eqref{eq:total_noise_correlation} can be written in the conventional form
\begin{equation}
    \left\langle
        \xi_i(t)\xi_j(t')
    \right\rangle
    =
    2m\Gamma_{\mathrm{phy}}\kb
    T_{\mathrm{bath}}^{\mathrm{COM}}\,
    \delta_{ij}\delta(t-t').
\end{equation}
When the internal--COM coupling is disabled, \(\Gamma_{\mathrm{int}}=0\), and the effective mechanical bath reduces to the background gas:
\begin{equation}
    T_{\mathrm{bath}}^{\mathrm{COM}}
    =
    T_{\mathrm{gas}}.
\end{equation}

The temperature \(T_{\mathrm{bath}}^{\mathrm{COM}}\) must not be confused with the instantaneous kinetic temperature calculated from the simulated velocity. The latter is a dynamical diagnostic that can contain secular oscillations, AC micromotion, transients, and statistical fluctuations. By contrast, \(T_{\mathrm{bath}}^{\mathrm{COM}}\) characterizes the effective reservoirs responsible for mechanical damping and diffusion.

In the numerical integration, the two stochastic contributions are combined into a single Gaussian velocity increment. For each particle, coordinate, and integration step,
\begin{equation}
    \Delta v_i^{\mathrm{th}}
    =
    \sqrt{
        \frac{2\kb}{m}
        \left(
            \Gamma_{\mathrm{gas}}T_{\mathrm{gas}}
            +
            \Gamma_{\mathrm{int}}T_{\mathrm{int}}
        \right)
        \Delta t
    }\,
    \eta_i,
    \label{eq:discrete_thermal_kick}
\end{equation}
where \(\Delta t\) is the integration time step and
\begin{equation}
    \eta_i\sim\mathcal{N}(0,1)
\end{equation}
is a normally distributed random number with zero mean and unit variance. A new independent value of \(\eta_i\) is generated for each coordinate, nanoparticle, and integration step. This discrete stochastic increment corresponds to an Euler--Maruyama treatment of the Langevin contribution \cite{kloeden1992,higham2001}.

Although Eq.~\eqref{eq:total_random_force} separates the gas and internal contributions conceptually, they do not need to be sampled separately in the numerical implementation. The sum of two independent Gaussian random forces is statistically equivalent to a single Gaussian force with the combined variance appearing in Eq.~\eqref{eq:discrete_thermal_kick}.

For simulations containing multiple nanoparticles, the stochastic increments are generated independently for each particle. The particles can nevertheless develop correlated motion through Coulomb interactions and the common trapping potential. Thus, the noise sources are independent, but the resulting trajectories do not necessarily remain statistically independent.

In the absence of external driving, feedback, and time-dependent energy injection, the balance between physical damping and stochastic diffusion leads to
\begin{equation}
    \frac{1}{2}m\left\langle v_i^2\right\rangle
    =
    \frac{1}{2}\kb T_{\mathrm{bath}}^{\mathrm{COM}}
\end{equation}
for each Cartesian direction. In three dimensions,
\begin{equation}
    \frac{1}{2}m
    \left\langle
        v_x^2+v_y^2+v_z^2
    \right\rangle
    =
    \frac{3}{2}\kb T_{\mathrm{bath}}^{\mathrm{COM}}.
\end{equation}
These relations refer to ensemble or sufficiently long time averages. They should not be expected to hold instantaneously, particularly when the motion contains secular oscillations, AC micromotion, transients, or active feedback. If \(T_{\mathrm{int}}\) changes significantly with time, then \(T_{\mathrm{bath}}^{\mathrm{COM}}\) also becomes time dependent and the center-of-mass motion need not reach a stationary thermal state.

The feedback force is modeled separately as an ideal velocity-dependent force,
\begin{equation}
    F_{\mathrm{fb},i}
    =
    -m\Gamma_{\mathrm{fb},i}v_i,
    \label{eq:feedback_force_langevin}
\end{equation}
where \(\Gamma_{\mathrm{fb},i}\) is the feedback damping rate along direction \(i\). This force is deterministic and is not treated as a thermal reservoir. Consequently, \(\Gamma_{\mathrm{fb},i}\) does not contribute to the Langevin-noise amplitude in Eq.~\eqref{eq:discrete_thermal_kick}.

For a stable harmonic mode with approximately constant reservoir temperatures, the ideal stationary center-of-mass temperature is
\begin{equation}
    T_{\mathrm{COM},i}^{\mathrm{ss}}
    \simeq
    \frac{
        \Gamma_{\mathrm{gas}}T_{\mathrm{gas}}
        +
        \Gamma_{\mathrm{int}}T_{\mathrm{int}}
    }{
        \Gamma_{\mathrm{gas}}
        +
        \Gamma_{\mathrm{int}}
        +
        \Gamma_{\mathrm{fb},i}
    }.
    \label{eq:ideal_feedback_temperature}
\end{equation}
The superscript \(\mathrm{ss}\) denotes the steady state. When feedback is disabled, this expression reduces to \(T_{\mathrm{bath}}^{\mathrm{COM}}\). When \(\Gamma_{\mathrm{int}}=0\), it reduces to the single-gas-reservoir result
\begin{equation}
    T_{\mathrm{COM},i}^{\mathrm{ss}}
    \simeq
    \frac{
        \Gamma_{\mathrm{gas}}
    }{
        \Gamma_{\mathrm{gas}}
        +
        \Gamma_{\mathrm{fb},i}
    }
    T_{\mathrm{gas}}.
\end{equation}
These expressions are idealized stationary results. They are not applicable during strong transients, when \(T_{\mathrm{int}}\) changes rapidly, or when excessively high feedback damping destabilizes the numerical dynamics and produces energy injection.

The internal--COM coupling is phenomenological and effectively one-way at the level of the thermal model: \(T_{\mathrm{int}}\) determines the temperature of the additional mechanical reservoir, but the corresponding energy exchanged with the center-of-mass motion is not explicitly fed back into the internal-energy equation. It should therefore be interpreted as an effective coupling mechanism rather than a complete microscopic model of bidirectional energy exchange.

The current Langevin model consequently provides an efficient description of Brownian motion produced by the gas together with the optional phenomenological influence of \(T_{\mathrm{int}}\) on the center-of-mass dynamics. At very low pressures, where \(\Gamma_{\mathrm{gas}}\) and the associated \(\Gamma_{\mathrm{int}}\) become weak, neglected mechanisms such as photon recoil, electrical noise, and measurement backaction may become comparatively important. These limitations should be considered when comparing the simulations quantitatively with experiments performed in the ultrahigh-vacuum regime.

\subsection{Instantaneous kinetic temperature}

The instantaneous kinetic temperature associated with the center-of-mass motion is estimated from the translational kinetic energy \cite{dani2021},
\begin{equation}
  T_{\mathrm{kin}}
  =
  \frac{m(v_x^2+v_y^2+v_z^2)}{3\kb}.
\end{equation}
This expression follows from classical equipartition for three translational degrees of freedom. If the motion were in thermal equilibrium with a bath at temperature $T$, one would expect
\begin{equation}
  \left\langle
  \frac{1}{2}m(v_x^2+v_y^2+v_z^2)
  \right\rangle
  =
  \frac{3}{2}\kb T.
\end{equation}
Inverting this relation gives the kinetic-temperature estimator used by the simulator.

In the current implementation, the initial center-of-mass temperature parameter, $T_{\mathrm{COM}}(0)$, is used to initialize the particle velocities through a Maxwell--Boltzmann distribution,
\begin{equation}
  v_i(0) \sim \mathcal{N}
  \left(
  0,
  \frac{\kb T_{\mathrm{COM}}(0)}{m}
  \right),
  \qquad i=x,y,z.
\end{equation}
Therefore, the initial displayed value of $\Tcom$ is not imposed independently of the velocity state. Instead, it is initialized from the kinetic energy of the sampled initial velocities,
\begin{equation}
  \Tcom(0)
  =
  T_{\mathrm{kin}}(0).
\end{equation}
For a single stochastic trajectory, this sampled value can differ from the nominal input value $T_{\mathrm{COM}}(0)$, while the ensemble average is consistent with the requested initial temperature.

It is important to note that $T_{\mathrm{kin}}$ can fluctuate strongly, especially in a single stochastic trajectory. It depends on the instantaneous velocity of the particle and can therefore vary rapidly due to coherent trap oscillations, Brownian noise, feedback transients, or abrupt changes in parameters. For this reason, the simulator does not display only $T_{\mathrm{kin}}$ directly, but rather a smoothed diagnostic quantity identified as $\Tcom$.

\subsection{Smoothed evolution of \texorpdfstring{$\Tcom$}{TCOM}}

The center-of-mass temperature displayed by the simulator, $\Tcom$, is a relaxation average of the instantaneous kinetic temperature \cite{dani2021},
\begin{equation}
  \frac{d\Tcom}{dt}
  \approx
  \gamma_{\mathrm{smooth}}
  \left(T_{\mathrm{kin}}-\Tcom\right).
\end{equation}
This equation acts as a low-pass filter applied to the kinetic temperature. When $T_{\mathrm{kin}}$ changes rapidly, $\Tcom$ follows the average kinetic-energy scale but does not reproduce all instantaneous fluctuations.

In the updated implementation, the smoothing rate is no longer determined only by the gas damping or by the internal--COM coupling. Instead, it also accounts for the damping effectively applied by manual damping and feedback cooling. The rate used by the simulator is
\begin{equation}
  \gamma_{\mathrm{smooth}}
  =
  \max
  \left[
  \gamma_{\mathrm{diag}},
  2\Gamma_{\mathrm{COM}}^{\mathrm{appl}},
  10^{-6}\,\mathrm{s}^{-1}
  \right],
\end{equation}
where
\begin{equation}
  \Gamma_{\mathrm{COM}}^{\mathrm{appl}}
  =
  \Gamma_{\mathrm{phys}}
  +
  \Gamma_{\mathrm{manual}}^{\mathrm{appl}}
  +
  \Gamma_{\mathrm{fb}}^{\mathrm{appl}}.
\end{equation}
Here
\begin{equation}
  \Gamma_{\mathrm{phys}}
  =
  \Gamma_{\mathrm{gas}}
  +
  \Gamma_{\mathrm{int}},
\end{equation}
is the physical coupling rate of the center-of-mass motion to the gas and to the phenomenological internal bath. The terms
$\Gamma_{\mathrm{manual}}^{\mathrm{appl}}$ and
$\Gamma_{\mathrm{fb}}^{\mathrm{appl}}$ are the effective viscous damping rates inferred from the actually applied manual and feedback forces.

The additional diagnostic floor is chosen as
\begin{equation}
  \gamma_{\mathrm{diag}}
  =
  \max
  \left[
  20\,\mathrm{s}^{-1},
  0.02\,\omega_{\mathrm{sec}}
  \right],
\end{equation}
with
\begin{equation}
  \omega_{\mathrm{sec}}
  =
  \max(\omega_r,\omega_z).
\end{equation}
This pressure-independent floor prevents the displayed temperature from becoming artificially frozen in the ultrahigh-vacuum limit.

\subsection{Energy-balance equation}
\label{sec:dinamica_of_temp}

The internal temperature $\Tint$ represents the thermal state of the nanoparticle's internal degrees of freedom \cite{seletskiy2016,hebestreit2018}. Microscopically, this internal energy is mainly stored in lattice vibrations, or phonons, and can be modified by several channels: heat exchange with the surrounding gas, thermal radiation, parasitic optical absorption, resonant absorption by active ions, and energy extraction through anti-Stokes fluorescence \cite{pringsheim1929,landau1946,sheik2009,seletskiy2016}. In the simulator, these many microscopic degrees of freedom are not treated individually. Instead, they are condensed into a single effective thermal variable, $\Tint(t)$.

This corresponds to a lumped thermal model. The nanoparticle is assumed to have a well-defined internal temperature at each instant, meaning that internal thermalization is taken to be fast compared with the time scale over which $\Tint$ changes appreciably. This does not imply global thermal equilibrium. The gas, the radiation field, the optical field, and the center-of-mass motion may all be out of equilibrium with one another. The assumption is only that the internal degrees of freedom of the nanoparticle can be described by a single effective temperature.

The thermal evolution is written as an energy-balance equation,
\begin{equation}
  \Cth\frac{d\Tint}{dt}
  =
  \dot Q_{\mathrm{gas}}
  +
  \dot Q_{\mathrm{rad}}
  +
  \dot Q_{\mathrm{laser}}.
  \label{eq:energy_balance_internal}
\end{equation}
Here $\Cth$ is the effective heat capacity of the nanoparticle. In the simplest case of a homogeneous spherical particle, it is proportional to the particle volume, density, and specific heat. In the more general core--shell case, it represents the sum of the heat capacities of the core and shell contributions. Thus, $\Cth$ sets the thermal inertia of the particle: for a given net power, a larger $\Cth$ produces a slower temperature change, whereas a smaller $\Cth$ makes the internal temperature respond more rapidly.

The sign convention is defined in terms of power delivered to the internal degrees of freedom:
\begin{equation}
  \dot Q>0
  \quad\Rightarrow\quad
  \text{heating of the nanoparticle},
\end{equation}
and
\begin{equation}
  \dot Q<0
  \quad\Rightarrow\quad
  \text{cooling of the nanoparticle}.
\end{equation}
With this convention, $\dot Q_{\mathrm{gas}}$ is positive when the gas transfers heat to the particle and negative when the gas removes heat from it. Similarly, $\dot Q_{\mathrm{rad}}$ is positive when the radiative environment heats the particle and negative when the particle loses energy by thermal emission. The optical term $\dot Q_{\mathrm{laser}}$ may be either positive or negative, depending on the balance between parasitic absorption, resonant absorption, fluorescence, and anti-Stokes cooling.

In the numerical implementation, the gas and radiative contributions are naturally computed as thermal powers and are then divided by $\Cth$. The laser contribution is implemented directly as a temperature rate,
\begin{equation}
  \frac{d\Tint}{dt}
  =
  \frac{\dot Q_{\mathrm{gas}}}{\Cth}
  +
  \frac{\dot Q_{\mathrm{rad}}}{\Cth}
  +
  \left(\frac{d\Tint}{dt}\right)_{\mathrm{laser}}.
  \label{eq:temp_int}
\end{equation}
The two forms are equivalent if
\begin{equation}
  \left(\frac{d\Tint}{dt}\right)_{\mathrm{laser}}
  =
  \frac{\dot Q_{\mathrm{laser}}}{\Cth}.
\end{equation}
This notation is convenient because the optical term can be evaluated from the active laser beams, local optical intensity, absorption spectrum, emission spectrum, parasitic loss, and cooling efficiency. In contrast, the gas and radiation channels are more transparently expressed as heat-transfer powers.

Equation~\eqref{eq:temp_int} describes the competition between thermalization and optical heating or cooling. The gas contribution tends to drive $\Tint$ toward the gas temperature $\Tgas$. The radiative contribution tends to drive $\Tint$ toward the effective radiative temperature of the environment, $\Trad$. The laser contribution can heat or cool the particle. It heats the particle when parasitic absorption or Stokes-shifted processes dominate, and it cools the particle when anti-Stokes fluorescence removes more energy than is deposited by absorption.

A thermal steady state is reached when the net power vanishes,
\begin{equation}
  \dot Q_{\mathrm{gas}}
  +
  \dot Q_{\mathrm{rad}}
  +
  \dot Q_{\mathrm{laser}}
  =
  0.
  \label{eq:thermal_steady_state}
\end{equation}
At this point, the internal temperature is constant, but it does not necessarily equal $\Tgas$ or $\Trad$. If optical absorption dominates, the steady-state internal temperature can lie above the environment temperature. If anti-Stokes laser cooling dominates, the particle can reach a steady state below the gas temperature, provided that gas heating, radiative exchange, and parasitic absorption do not compensate the optical cooling power.

Therefore, $\Tint$ should be interpreted as the result of a dynamic energy balance rather than as a prescribed bath temperature. It is the internal thermal state selected by the simultaneous action of gas exchange, radiative exchange, and laser-induced heating or cooling.

\subsection{Relation between \texorpdfstring{$\Tint$}{Tint} and \texorpdfstring{$\Tcom$}{TCOM}}

The distinction between $\Tint$ and $\Tcom$ is essential for correctly interpreting the simulator results \cite{hebestreit2018,dani2021}. The internal temperature $\Tint$ measures the energy stored in the nanoparticle material in the form of phonons, heat deposited by optical absorption, energy removed by anti-Stokes emission, and thermal exchange with gas/radiation. By contrast, $\Tcom$ measures the translational mechanical energy of the center of mass.

These two temperatures can evolve very differently. For example, parasitic light absorption can strongly increase $\Tint$ without immediately heating the center of mass. Conversely, gas noise or trap instability can increase $\Tcom$ even when $\Tint$ remains close to ambient temperature. The phenomenological coupling $\GammaInt$ is the mechanism that allows the simulator to explore an indirect influence of $\Tint$ on $\Tcom$.

When the internal--COM coupling is off, the internal temperature does not participate in the amplitude of the mechanical noise \cite{kloeden1992,dani2021}. When the coupling is on, the center of mass feels an effective bath combining gas and internal temperature.

This construction is useful for investigating scenarios in which internal optical heating can compromise mechanical stability. For example, if $\Tint$ increases due to parasitic absorption and $\GammaInt$ is significant, the mechanical noise increases, which can raise $\Tcom$ and enlarge the oscillation amplitude of the particle. Conversely, if a laser-cooling process reduces $\Tint$ below $\Tgas$, the same phenomenological channel can represent a tendency to reduce the effective mechanical noise. This can be a mechanism for improving the ground-state fidelity of the center of mass in microscopic systems.

The coupling $\GammaInt$ is deliberately phenomenological. It does not replace a complete microscopic model of energy transfer between internal and translational degrees of freedom. Its role in the simulator is to allow a controlled exploration of the question: ``what happens to the mechanical dynamics if the internal temperature also contributes to the effective center-of-mass bath?'' Therefore, quantitative results involving $\GammaInt$ should be interpreted as model trends, not as absolute material predictions without experimental calibration.

\section{Thermal exchange with the gas}

Thermal exchange with the gas describes the transfer of energy between environmental gas molecules and the internal degrees of freedom of the nanoparticle \cite{epstein1924,hebestreit2018}. In the simulator, this mechanism acts on the internal temperature $\Tint$ and represents the tendency of gas collisions to drive the particle toward the gas temperature $\Tgas$. This channel is especially relevant at moderate and high pressures, where frequent molecular collisions provide an efficient thermal link. In high vacuum, the molecular density is reduced, collisions become rare, and gas-mediated heat exchange can become weak compared with optical heating or cooling and thermal radiation.

The simulator contains two descriptions of internal gas exchange. The main physical description is based on a heat-transfer power, whose form depends on the Knudsen regime. This is the preferred option for quantitative studies of the internal temperature because it treats gas exchange as a thermal power entering the internal energy balance. The simulator also retains a legacy phenomenological mode, in which the internal temperature relaxes toward $\Tgas$ with a rate proportional to the mechanical gas damping. This second option is useful for simple qualitative tests, but it should not be confused with a microscopic or kinetic heat-transfer model.

\subsection{Knudsen number and transport regimes}

The gas transport regime is determined by the Knudsen number \cite{epstein1924,hebestreit2018},
\begin{equation}
  \mathrm{Kn}
  =
  \frac{\lambda_{\mathrm{mfp}}}{2R},
\end{equation}
where $\lambda_{\mathrm{mfp}}$ is the mean free path of the gas molecules and $2R$ is the nanoparticle diameter. The mean free path is estimated as
\begin{equation}
  \lambda_{\mathrm{mfp}}
  =
  \frac{\kb \Tgas}{\sqrt{2}\pi d_{\mathrm{air}}^2P},
\end{equation}
where $d_{\mathrm{air}}$ is an effective molecular diameter and $P$ is the gas pressure. The inverse dependence on pressure is central. As the pressure is reduced, gas molecules travel longer distances before colliding with each other, and heat transfer around the particle gradually changes from continuum conduction to molecular transport.

When $\mathrm{Kn}\ll1$, the mean free path is much smaller than the particle diameter. Many gas--gas collisions occur near the particle surface, and heat exchange can be described approximately as thermal diffusion in a continuous medium. When $\mathrm{Kn}\sim1$, the system lies in the transition or intermediate regime, where neither the continuum picture nor the purely free-molecular picture is fully adequate. When $\mathrm{Kn}\gg1$, the mean free path is much larger than the particle size, so molecules reach the particle almost independently and exchange energy with the surface through individual collisions. The simulator uses the Knudsen number to select the appropriate effective thermal conductance between the particle and the gas.

\subsection{Intermediate Knudsen regime}

In the intermediate regime, the simulator writes the gas heat flow as a linear conductance law,
\begin{equation}
  \dot Q_{\mathrm{gas}}^{\mathrm{int}}
  =
  -G_{\mathrm{gas}}
  \left(\Tint-\Tgas\right).
  \label{eq:qgas_intermediate}
\end{equation}
The sign convention is the same as in the internal energy-balance equation, positive power heats the particle, while negative power cools it. Therefore, if $\Tint>\Tgas$, the factor $\Tint-\Tgas$ is positive and Eq.~\eqref{eq:qgas_intermediate} gives $\dot Q_{\mathrm{gas}}^{\mathrm{int}}<0$, meaning that the gas removes heat from the particle. If $\Tint<\Tgas$, the gas transfers heat to the particle.

The thermal conductance used in this branch is
\begin{equation}
  G_{\mathrm{gas}}
  =
  \frac{8\pi R^2 k_{\mathrm{air}}}{2R+\lambda_{\mathrm{mfp}}G},
  \label{eq:ggas_intermediate}
\end{equation}
with
\begin{equation}
  G
  =
  \frac{18\gamma_{\mathrm{sh}}-10}
  {a_{\mathrm{acc}}(\gamma_{\mathrm{sh}}+1)}.
  \label{eq:ggas_slip_factor}
\end{equation}
Here $k_{\mathrm{air}}$ is the thermal conductivity of the gas, $a_{\mathrm{acc}}$ is the thermal accommodation coefficient, and $\gamma_{\mathrm{sh}}$ is the heat-capacity ratio of the gas. The accommodation coefficient describes how efficiently a molecule exchanges thermal energy with the particle surface during a collision. If $a_{\mathrm{acc}}$ is close to unity, the outgoing molecules are assumed to be nearly thermalized with the surface. If $a_{\mathrm{acc}}$ is small, collisions are less effective and the thermal conductance is reduced.

Equation~\eqref{eq:ggas_intermediate} interpolates between a size-limited conduction picture and a mean-free-path-limited picture. When $\lambda_{\mathrm{mfp}}\ll2R$, the denominator is controlled mainly by the particle diameter, and the exchange resembles continuum heat conduction around a small object. When $\lambda_{\mathrm{mfp}}$ becomes comparable to or larger than the particle size, the term proportional to $\lambda_{\mathrm{mfp}}$ reduces the conductance. This expresses the physical fact that, at lower pressure, fewer molecules participate in carrying heat between the particle and the gas.

\subsection{Free-molecular regime}

The free-molecular regime applies when $\mathrm{Kn}\gg1$, so that the gas mean free path is much larger than the particle diameter. In this limit, the usual continuum picture of a smooth gas temperature field around the particle is no longer appropriate. The gas molecules travel essentially ballistically between collisions, and the heat transfer is better described by the flux of molecules that reaches the particle surface.

The gas-mediated heat flow is written as
\begin{equation}
  \dot Q_{\mathrm{gas}}^{\mathrm{Kn}}
  =
  -G_{\mathrm{Kn}}
  \left(\Tint-\Tgas\right),
  \label{eq:qgas_kn_conductance}
\end{equation}
where $G_{\mathrm{Kn}}$ is the free-molecular thermal conductance. This equation has the same sign structure as the intermediate-regime expression. The gas cools the particle when $\Tint>\Tgas$ and heats it when $\Tint<\Tgas$, but the strength of this exchange is now controlled by the molecular collision rate rather than by continuum conduction.

The simulator uses the kinetic-theory-inspired conductance
\begin{equation}
  G_{\mathrm{Kn}}
  =
  a_{\mathrm{acc}}
  \sqrt{\frac{2}{3\pi}}\,
  \pi R^2\,
  \bar v_g\,
  \frac{\gamma_g-1}{\gamma_g+1}\,
  \frac{P}{\Tgas},
  \label{eq:gkn}
\end{equation}
where $\gamma_g$ is the gas heat-capacity ratio. Combining Eqs.~\eqref{eq:qgas_kn_conductance} and \eqref{eq:gkn}, the free-molecular heat flow can be written as
\begin{equation}
  \dot Q_{\mathrm{gas}}^{\mathrm{Kn}}
  =
  -a_{\mathrm{acc}}
  \sqrt{\frac{2}{3\pi}}\,
  \pi R^2\,
  \bar v_g\,
  \frac{\gamma_g-1}{\gamma_g+1}\,
  P
  \left(
  \frac{\Tint}{\Tgas}-1
  \right).
  \label{eq:qgas_knudsen}
\end{equation}

This expression separates the main physical ingredients of molecular heat exchange. The factor $\pi R^2$ represents the geometrical collision cross section of the particle. The pressure $P$ controls the molecular density and therefore the number of molecules reaching the particle per unit time. The mean speed $\bar v_g$ sets the characteristic molecular flux. The accommodation coefficient $a_{\mathrm{acc}}$ determines how effectively the molecules exchange thermal energy with the surface. The factor involving $\gamma_g$ accounts phenomenologically for the thermodynamic properties of the gas. Finally, the term $(\Tint/\Tgas-1)$ measures the thermal imbalance between the nanoparticle and the gas bath.

The defining feature of this regime is the approximate proportionality
\begin{equation}
  G_{\mathrm{Kn}}\propto P.
\end{equation}
Reducing the pressure directly reduces the number of gas molecules that collide with the particle per unit time. Consequently, the gas becomes a weaker thermal reservoir for the internal temperature. At sufficiently low pressure, $\Tint$ is no longer strongly clamped to $\Tgas$ and can be governed mainly by the balance between laser-induced heating or cooling and thermal radiation. This is why, in ultrahigh vacuum, even modest optical absorption may noticeably heat the particle, while efficient anti-Stokes fluorescence may cool it below the gas temperature if the remaining thermal links do not compensate the optical cooling power.

\subsection{Legacy internal--gas mode}

The simulator also retains a legacy internal--gas mode in which the heat exchange is represented as a direct relaxation of $\Tint$ toward $\Tgas$,
\begin{equation}
  \dot Q_{\mathrm{gas}}^{\mathrm{legacy}}
  =
  -\gamma_{\mathrm{int-gas}}\Cth
  \left(\Tint-\Tgas\right),
  \label{eq:qgas_legacy}
\end{equation}
with
\begin{equation}
  \gamma_{\mathrm{int-gas}}
  =
  \chi_{\mathrm{int-gas}}\GammaGas.
  \label{eq:gamma_int_gas_legacy}
\end{equation}
Here $\chi_{\mathrm{int-gas}}$ is an adjustable dimensionless factor and $\GammaGas$ is the mechanical gas damping rate. This mode is simple and numerically convenient because it imposes a tunable thermal relaxation time for the internal temperature. However, it ties internal heat exchange to center-of-mass damping, which are physically distinct processes. Mechanical damping describes momentum dissipation of the external motion, whereas internal gas exchange describes energy transfer between gas molecules and the thermal degrees of freedom of the particle.

For this reason, the legacy mode should be interpreted as a phenomenological control rather than as a quantitative thermal model. For studies in which the absolute internal temperature, pressure dependence, or competition with optical refrigeration is important, the Knudsen-based power mode is the more physically coherent choice.

\section{Thermal radiation}

Thermal radiation is the channel through which the nanoparticle exchanges energy with the thermal electromagnetic field of its surroundings \cite{bohren1983,hebestreit2018}. Unlike gas-mediated heat exchange, radiative exchange does not require molecular collisions. Therefore, even in high vacuum, where the gas thermal conductance can become very small, the particle can still emit and absorb infrared radiation. This makes thermal radiation an important contribution to the internal energy balance, especially at low pressures or when the internal temperature becomes significantly different from the chamber-wall temperature.

In the simulator, thermal radiation contributes to the internal temperature equation through a net radiative power,
\begin{equation}
  \Cth\frac{d\Tint}{dt}
  =
  \cdots
  +
  \dot Q_{\mathrm{rad}}
  +
  \cdots ,
\end{equation}
with the same sign convention used for the gas term. A positive value of $\dot Q_{\mathrm{rad}}$ means that radiation delivers net energy to the internal degrees of freedom of the particle, increasing $\Tint$. A negative value means that the particle emits more radiative power than it absorbs from the environment, decreasing $\Tint$.

The simulator allows different levels of description for this radiative channel. The simplest one is a gray-body Stefan--Boltzmann model, in which the complicated spectral response of the particle is represented by a single effective emissivity. More detailed spectral modes use an absorption cross section, either in the Rayleigh approximation or from a finite-size Mie calculation, and integrate this response over the thermal Planck spectrum. The gray-body model is therefore fast and phenomenological, whereas the spectral models are more closely connected to the wavelength-dependent infrared absorption of the particle.

\subsection{Gray-body model}

In gray-body mode, the nanoparticle is treated as a small object with an effective emissivity $\epsilon_{\mathrm{bb}}$ and surface area $4\pi R^2$. The net radiative power is written as
\begin{equation}
  \dot Q_{\mathrm{rad}}^{\mathrm{gray}}
  =
  4\pi R^2
  \epsilon_{\mathrm{bb}}
  \sigma_{\mathrm{SB}}
  \left(
  \Trad^4-\Tint^4
  \right),
  \label{eq:qrad_gray}
\end{equation}
where $\sigma_{\mathrm{SB}}$ is the Stefan--Boltzmann constant and $\Trad$ is the effective radiative temperature of the environment, such as the chamber-wall temperature.

This expression is the difference between the power absorbed from a gray thermal environment at temperature $\Trad$ and the power emitted by the particle at temperature $\Tint$. If $\Tint>\Trad$, then $\Trad^4-\Tint^4<0$, and Eq.~\eqref{eq:qrad_gray} gives $\dot Q_{\mathrm{rad}}^{\mathrm{gray}}<0$. In this case, radiation removes internal energy from the nanoparticle. If $\Tint<\Trad$, the particle absorbs more thermal radiation from the environment than it emits, and the radiative term heats the particle.

The parameter $\epsilon_{\mathrm{bb}}$ is an effective gray-body emissivity. It does not resolve any specific infrared band, resonance, or material dispersion. Instead, it summarizes the overall ability of the particle to emit and absorb thermal radiation. Values close to zero suppress radiative exchange, while larger values make the particle behave more like an efficient thermal emitter. Since the gray-body power scales as $T^4$, this channel becomes increasingly important when the internal temperature rises.

The gray-body model is useful when one needs a robust and computationally inexpensive representation of radiative exchange without specifying detailed infrared optical constants. It is less appropriate when the particle is small compared with the relevant thermal wavelengths, when the material is nearly transparent in the infrared, or when specific absorption bands dominate the radiative cooling or heating.

\subsection{Spectral Rayleigh and Mie models}

In the spectral radiation modes, the simulator computes radiative exchange by integrating the particle absorption cross section over the blackbody spectrum. This implements Kirchhoff's law in a form appropriate for a small particle, where the same optical response that determines absorption at a given wavelength also determines emission at that wavelength \cite{bohren1983}. The emitted and absorbed powers are written as
\begin{equation}
  P_{\mathrm{emit}}(\Tint)
  =
  4\int_{\lambda_{\min}}^{\lambda_{\max}}
  C_{\mathrm{abs}}(\lambda,R)\,
  M_\lambda(\lambda,\Tint)\,
  d\lambda,
  \label{eq:pemit_spectral}
\end{equation}
and
\begin{equation}
  P_{\mathrm{abs,env}}(\Trad)
  =
  4\int_{\lambda_{\min}}^{\lambda_{\max}}
  C_{\mathrm{abs}}(\lambda,R)\,
  M_\lambda(\lambda,\Trad)\,
  d\lambda.
  \label{eq:pabsenv_spectral}
\end{equation}
The net radiative power is then
\begin{equation}
  \dot Q_{\mathrm{rad}}^{\mathrm{spec}}
  =
  P_{\mathrm{abs,env}}(\Trad)
  -
  P_{\mathrm{emit}}(\Tint).
  \label{eq:qrad_spectral}
\end{equation}
The factor of four appears because $M_\lambda$ is the hemispherical blackbody spectral exitance, while $C_{\mathrm{abs}}$ is an absorption cross section for incident radiation. Equivalently, this form is consistent with the standard relation between blackbody radiance and exitance for an isotropic thermal radiation field.

The blackbody spectral exitance is
\begin{equation}
  M_\lambda(\lambda,T)
  =
  \frac{2\pi h c^2}{\lambda^5}
  \frac{1}
  {
  \exp\left(
  \frac{hc}{\lambda\kb T}
  \right)-1
  }.
  \label{eq:planck_spectral_exitance}
\end{equation}
It gives the emitted blackbody power per unit surface area and per unit wavelength. The absorption cross section $C_{\mathrm{abs}}(\lambda,R)$ then selects which parts of the thermal spectrum are actually relevant for the nanoparticle. As a result, the spectral radiation model can represent situations in which a hot particle emits weakly because its infrared absorption is small, or emits efficiently only in certain wavelength bands.

\subsubsection{Rayleigh spectral mode}
\label{sec:rayleigh_spectral_mode}

In the Rayleigh spectral mode, the nanoparticle is treated as a homogeneous and isotropic sphere in vacuum whose radius is small compared with the thermally relevant wavelengths. The approximation is characterized by the size parameter
\begin{equation}
  x(\lambda)
  =
  \frac{2\pi R}{\lambda},
  \label{eq:rayleigh_size_parameter}
\end{equation}
which must satisfy $x\ll 1$ over the wavelength interval that contributes significantly to the radiative exchange. The more restrictive condition
\begin{equation}
  \left|\widetilde{m}(\lambda)\right|x(\lambda)\ll 1,
\end{equation}
where $\widetilde{m}$ is the complex refractive index of the particle relative to its surroundings, may also be used. Under these conditions, retardation inside the particle is negligible and the optical response is dominated by an induced electric dipole.

For a sphere of radius $R$ in vacuum, the quasistatic electric polarizability is \cite{bohren1983}
\begin{equation}
  \alpha_{\mathrm{qs}}(\lambda,R)
  =
  4\pi\varepsilon_0R^3
  \frac{\varepsilon(\lambda)-1}
       {\varepsilon(\lambda)+2},
  \label{eq:rayleigh_quasistatic_polarizability}
\end{equation}
where $\varepsilon(\lambda)$ is the complex relative dielectric function of the particle. The leading-order dipolar absorption cross section used by the simulator is
\begin{equation}
  C_{\mathrm{abs}}^{\mathrm{Ray}}(\lambda,R)
  =
  \frac{k}{\varepsilon_0}
  \operatorname{Im}
  \left[
  \alpha_{\mathrm{qs}}(\lambda,R)
  \right],
  \label{eq:rayleigh_cabs_from_polarizability}
\end{equation}
with $k=2\pi/\lambda$. Consequently,
\begin{equation}
  \boxed{
  C_{\mathrm{abs}}^{\mathrm{Ray}}(\lambda,R)
  =
  4\pi kR^3
  \operatorname{Im}
  \left[
  \frac{\varepsilon(\lambda)-1}
       {\varepsilon(\lambda)+2}
  \right]
  }.
  \label{eq:rayleigh_absorption_cross_section}
\end{equation}

The complex refractive index and dielectric function are related by
\begin{equation}
  \widetilde{m}(\lambda)
  =
  n(\lambda)+i\kappa(\lambda),
  \label{eq:rayleigh_complex_refractive_index}
\end{equation}
and
\begin{equation}
  \varepsilon(\lambda)
  =
  \widetilde{m}^{\,2}(\lambda)
  =
  \varepsilon'(\lambda)+i\varepsilon''(\lambda),
  \label{eq:rayleigh_dielectric_function}
\end{equation}
where
\begin{align}
  \varepsilon'(\lambda)
  &=
  n^2(\lambda)-\kappa^2(\lambda),
  \\
  \varepsilon''(\lambda)
  &=
  2n(\lambda)\kappa(\lambda).
\end{align}
For a particle in vacuum,
\begin{equation}
  \operatorname{Im}
  \left[
  \frac{\varepsilon-1}{\varepsilon+2}
  \right]
  =
  \frac{3\varepsilon''}
  {\left(\varepsilon'+2\right)^2+\left(\varepsilon''\right)^2}.
  \label{eq:rayleigh_polarizability_imaginary_part}
\end{equation}
Thus, radiative coupling is governed by the absorptive dielectric response rather than by geometrical area alone. A transparent material with a very small extinction coefficient $\kappa$ can exchange much less thermal radiation than a gray-body model with constant emissivity would predict.

In particular,
\begin{equation}
  C_{\mathrm{abs}}^{\mathrm{Ray}}
  \propto R^3,
  \label{eq:rayleigh_radius_scaling}
\end{equation}
whereas the gray-body Stefan--Boltzmann model scales with the surface area $A=4\pi R^2$. The Rayleigh response is therefore proportional to the particle volume multiplied by its absorptive dielectric contrast, rather than to independent emission from elements of its geometrical surface.

In the current implementation, $n(\lambda)$ and $\kappa(\lambda)$ are obtained from approximate built-in infrared models containing a background extinction coefficient and broad resonances associated with the selected material preset. The user-controlled infrared absorption scale $s_{\mathrm{IR}}$ modifies the extinction coefficient according to
\begin{equation}
  \kappa(\lambda)
  \longrightarrow
  s_{\mathrm{IR}}\kappa(\lambda).
  \label{eq:rayleigh_ir_scale}
\end{equation}
The simulator then recalculates $\varepsilon(\lambda)$ and $C_{\mathrm{abs}}^{\mathrm{Ray}}$. Because $\kappa$ enters both the numerator and denominator of Eq.~\eqref{eq:rayleigh_polarizability_imaginary_part}, $s_{\mathrm{IR}}$ is not, in general, a direct multiplicative factor on the cross section; approximate linearity is recovered only in the weak-absorption limit. These optical constants are phenomenological and should not be interpreted as a complete experimentally measured dielectric function.

The spectra supplied through the laser-cooling CSV interface are not used automatically to construct this infrared dielectric function. The active-ion laser spectrum and the broadband thermal-radiation model are treated as distinct optical channels, avoiding the interpretation of near-infrared electronic transitions as the full infrared response of the host material.

Because Eq.~\eqref{eq:rayleigh_absorption_cross_section} factorizes as
\begin{equation}
  C_{\mathrm{abs}}^{\mathrm{Ray}}(\lambda,R)
  =
  R^3\mathcal{C}_{\mathrm{Ray}}(\lambda),
\end{equation}
the simulator precomputes
\begin{equation}
  \frac{P_{\mathrm{Ray}}(T,R)}{R^3}
  =
  4
  \int_{\lambda_{\min}}^{\lambda_{\max}}
  \mathcal{C}_{\mathrm{Ray}}(\lambda)
  M_\lambda(T)\,d\lambda.
  \label{eq:rayleigh_power_per_r3}
\end{equation}
During time integration, the interpolated value is multiplied by $R^3$, avoiding repeated evaluation of the full spectral integral. Numerically, the integral extends from $1$ to $100~\mu\mathrm{m}$ and uses 260 logarithmically spaced wavelength points with trapezoidal integration. The power is tabulated from approximately $1$ to $10\,000~\mathrm{K}$ using temperature steps of $5~\mathrm{K}$ at low temperatures, $25~\mathrm{K}$ at intermediate temperatures, and $100~\mathrm{K}$ at the highest temperatures; linear interpolation is used between tabulated values.

For diagnostic purposes, the result may be expressed as an effective temperature-dependent emissivity,
\begin{equation}
  \varepsilon_{\mathrm{eff}}(T,R)
  =
  \frac{P_{\mathrm{Ray}}(T,R)}
  {4\pi R^2\sigma_{\mathrm{SB}}T^4}.
  \label{eq:rayleigh_effective_emissivity}
\end{equation}
This derived quantity is distinct from the constant gray-body emissivity, whose user input is inactive when the Rayleigh spectral mode is selected.

When Rayleigh mode is selected manually, the dipolar expression is applied throughout the integration interval even if $x$ becomes too large at some wavelengths. Its validity should therefore be assessed over the wavelengths that contribute appreciably to both $M_\lambda(T_{\mathrm{int}})$ and $M_\lambda(T_{\mathrm{rad}})$, rather than at the trapping- or cooling-laser wavelength alone.

In the automatic spectral mode, the simulator estimates a characteristic wavelength using Wien's displacement law,
\begin{equation}
  \lambda_{\mathrm{peak}}
  =
  \frac{b}{T_{\mathrm{ref}}},
  \qquad
  T_{\mathrm{ref}}
  =
  \max\left(T_{\mathrm{int}},T_{\mathrm{rad}}\right),
  \label{eq:auto_spectral_peak_wavelength}
\end{equation}
and evaluates
\begin{equation}
  x_{\mathrm{peak}}
  =
  \frac{2\pi R}{\lambda_{\mathrm{peak}}}.
\end{equation}
The Rayleigh model is used for $x_{\mathrm{peak}}<0.35$; otherwise, the simulator switches to the finite-size Mie calculation.

The present Rayleigh implementation retains only the leading electric-dipole absorption term. It excludes retardation, radiative-reaction corrections, magnetic-dipole contributions, and higher-order multipoles. Rayleigh scattering,
\begin{equation}
  C_{\mathrm{sca}}^{\mathrm{Ray}}
  =
  \frac{8\pi}{3}
  k^4R^6
  \left|
  \frac{\varepsilon-1}{\varepsilon+2}
  \right|^2,
\end{equation}
is also neglected because it is of higher order in $R$. When these finite-size or scattering effects become relevant, the Mie spectral mode is more appropriate.

Finally, core--shell particles are not described by a multilayer polarizability: the calculation uses the external radius and one effective set of optical constants. Temperature dependence, crystallographic anisotropy, surface effects, and size-dependent optical constants are likewise neglected. The quantitative accuracy of the Rayleigh mode is therefore ultimately limited by the quality of the infrared optical constants supplied to the calculation.

\subsubsection{Mie spectral mode}

The Mie spectral mode retains the homogeneous, isotropic sphere-in-vacuum model introduced above but removes the electric-dipole restriction. Its absorption cross section is
\begin{equation}
  C_{\mathrm{abs}}^{\mathrm{Mie}}(\lambda,R)
  =
  \pi R^2Q_{\mathrm{abs}}(\lambda,R),
  \label{eq:mie_absorption_cross_section}
\end{equation}
where the absorption efficiency is calculated internally as
\begin{equation}
  Q_{\mathrm{abs}}
  =
  Q_{\mathrm{ext}}-Q_{\mathrm{sca}}.
  \label{eq:mie_absorption_efficiency}
\end{equation}
For a sphere, the extinction and scattering efficiencies are
\begin{equation}
  Q_{\mathrm{ext}}
  =
  \frac{2}{x^2}
  \sum_{\ell=1}^{\ell_{\max}}
  (2\ell+1)
  \operatorname{Re}\left(a_\ell+b_\ell\right),
  \label{eq:mie_extinction_efficiency}
\end{equation}
and
\begin{equation}
  Q_{\mathrm{sca}}
  =
  \frac{2}{x^2}
  \sum_{\ell=1}^{\ell_{\max}}
  (2\ell+1)
  \left(|a_\ell|^2+|b_\ell|^2\right),
  \label{eq:mie_scattering_efficiency}
\end{equation}
where $a_\ell$ and $b_\ell$ are the electric and magnetic Mie coefficients. They depend on
\begin{equation}
  x
  =
  \frac{2\pi R}{\lambda},
  \label{eq:mie_size_parameter}
\end{equation}
and
\begin{equation}
  \widetilde{m}(\lambda)
  =
  n(\lambda)+i\kappa(\lambda),
  \label{eq:mie_complex_refractive_index}
\end{equation}
so $Q_{\mathrm{abs}}$ varies with wavelength, radius, and optical constants.

The Mie calculation uses the same built-in infrared optical models and the same transformation $\kappa\rightarrow s_{\mathrm{IR}}\kappa$ as the Rayleigh mode. After this transformation, the complete Mie efficiencies are recalculated, so $s_{\mathrm{IR}}$ is not applied directly as a multiplicative factor to $Q_{\mathrm{abs}}$.

The thermal power is obtained by integrating the Mie absorption cross section against the Planck spectrum,
\begin{equation}
  P_{\mathrm{Mie}}(T,R)
  =
  4
  \int_{\lambda_{\min}}^{\lambda_{\max}}
  C_{\mathrm{abs}}^{\mathrm{Mie}}(\lambda,R)
  M_\lambda(T)\,d\lambda,
  \label{eq:mie_thermal_power}
\end{equation}
with $M_\lambda(T)$ given by Eq.~\eqref{eq:planck_spectral_exitance}. The corresponding net radiative heat flow is
\begin{equation}
  Q_{\mathrm{rad}}
  =
  P_{\mathrm{Mie}}(T_{\mathrm{rad}},R)
  -
  P_{\mathrm{Mie}}(T_{\mathrm{int}},R),
  \label{eq:mie_net_radiative_power}
\end{equation}
with the same sign convention as in the Rayleigh mode.

The simulator evaluates Eq.~\eqref{eq:mie_thermal_power} on the same logarithmic wavelength grid from $1$ to $100~\mu\mathrm{m}$. The Mie series is truncated according to
\begin{equation}
  \ell_{\max}
  \simeq
  \left\lceil
  x+4x^{1/3}+8
  \right\rceil.
\end{equation}
For $x<0.08$, the Rayleigh result is used directly to avoid unnecessary evaluation of the full series.

Mie mode is relevant when the size parameter is no longer much smaller than unity over the thermally important wavelengths. It then accounts for retardation, scattering, internal-field interference, and higher-order multipoles omitted by the dipolar model. It nevertheless shares the material limitations described above: the particle is represented by its external radius and one effective set of optical constants, and no multilayer Mie calculation is performed for core--shell structures.

\section{Optical thermal term: absorption, emission, and laser cooling}

The optical thermal term describes how the optical fields modify the internal temperature of the nanoparticle \cite{ashkin1986,neuman2004,sheik2009,seletskiy2016}. In the simulator, this term is the part of the internal energy balance directly driven by the active optical tweezers. It combines two physically different mechanisms. The first is parasitic absorption, in which optical power is converted directly into heat through impurities, defects, off-resonant losses, or other nonradiative channels. The second is active absorption followed by fluorescence. Active absorption can either cool or heat the particle depending on whether the emitted photons carry, on average, more or less energy than the absorbed laser photons.

With the sign convention used in the internal energy equation, the laser contribution can be written as
\begin{equation}
  \dot Q_{\mathrm{laser}}
  =
  -\Pcool,
  \label{eq:q_laser_from_pcool}
\end{equation}
where $\Pcool>0$ means net optical cooling and $\Pcool<0$ means net optical heating. Equivalently, in the temperature equation,
\begin{equation}
  \left(\frac{d\Tint}{dt}\right)_{\mathrm{laser}}
  =
  -\frac{\Pcool}{\Cth}.
  \label{eq:dtint_laser_from_pcool}
\end{equation}
This convention makes the optical term consistent with the gas and radiation terms: positive thermal power heats the internal degrees of freedom, while positive cooling power lowers $\Tint$.

\subsection{Active and parasitic absorbed power}

For each optical tweezer, the simulator evaluates the local intensity $I$ at the particle position and uses it to estimate two absorbed powers. The active absorbed power, $\Pabsact$, is associated with the selected spectral transition or absorption band. The background absorbed power, $\Pabsbg$, represents parasitic loss and always enters as heating. The separation is essential because active absorption may participate in anti-Stokes laser cooling, whereas parasitic absorption deposits heat without producing useful fluorescence \cite{epstein1995,sheik2009,seletskiy2010,seletskiy2016,xia2021}.

In the optically thin approximation, the absorbed powers are
\begin{align}
  \Pabsact &= I\,\Vact\,\alpha_{\mathrm{active}}, \\
  \Pabsbg  &= I\,\Vtot\,\alpha_{\mathrm{bg}}.
  \label{eq:pabs_thin}
\end{align}
This approximation assumes that the optical depth is small, $\alpha\ell\ll1$, so that the absorbed fraction grows linearly with absorption coefficient, particle volume, and local intensity. The active volume $\Vact$ may represent the whole particle or only the optically active region, depending on whether the simulated geometry is homogeneous or core--shell. The total volume $\Vtot$ is used for background absorption because parasitic losses are treated as a bulk heating channel.

For larger optical depths, the simulator can use a finite-chord absorption model ,
\begin{align}
  \Pabsact
  &=
  I\,\Aact
  \left[
  1-\exp\left(-\alpha_{\mathrm{active}}\ell_{\mathrm{act}}\right)
  \right],
  \\
  \Pabsbg
  &=
  I\,\Atot
  \left[
  1-\exp\left(-\alpha_{\mathrm{bg}}\ell_{\mathrm{tot}}\right)
  \right].
  \label{eq:pabs_chord}
\end{align}
Here $\Aact$ and $\Atot$ are the effective projected areas used for the active and total regions, and the mean optical path through a spherical region is approximated by
\begin{equation}
  \ell \simeq \frac{4R}{3}.
\end{equation}
The factor $1-\exp(-\alpha\ell)$ is the absorbed fraction along this effective path. Unlike the optically thin expression, the finite-chord model saturates the absorbed fraction as the optical depth becomes large, preventing the absorbed power from increasing indefinitely in a purely linear way.

\subsection{Active absorption coefficient}

The active absorption coefficient is determined by the selected spectral source. In a generic preset, the simulator uses an internal phenomenological spectrum that is useful for qualitative exploration and visual validation. In a custom CSV spectrum, the absorption is supplied by the user and can represent a measured material. In generated-band mode, the absorption is constructed as a sum of adjustable phenomenological bands. These three sources differ in how the spectrum is generated, but they all provide the same quantity needed by the thermal model: an effective active absorption coefficient at the laser wavelength and current temperature.

If the spectral data provide an absorption cross section $\sigma_{\mathrm{abs}}$, the simulator converts it into a macroscopic absorption coefficient through
\begin{equation}
  \alpha_{\mathrm{active}}
  =
  N_{\mathrm{act}}\sigma_{\mathrm{abs}},
  \label{eq:alpha_from_sigma}
\end{equation}
where $N_{\mathrm{act}}$ is the effective density of active centers. This relation expresses the fact that the same microscopic cross section produces stronger macroscopic absorption when more active centers are present per unit volume. If the input data already provide $\alpha_{\mathrm{abs}}$ in $\mathrm{cm}^{-1}$, this value is used directly because it already includes the active-center density and represents the effective absorption of the material.

The magnitude of $\alpha_{\mathrm{active}}$ controls how strongly the laser couples to the active optical channel. If $\alpha_{\mathrm{active}}$ is very small, the active laser-cooling mechanism has little influence on $\Tint$. If $\alpha_{\mathrm{active}}$ is large, more active optical power is absorbed, but whether this absorption cools or heats the particle still depends on the fluorescence energy balance and on the external quantum efficiency.

\subsection{Absorption saturation}

When saturation is enabled, the active absorption coefficient is reduced according to a standard phenomenological saturation law \cite{mccumber1964,sheik2009},
\begin{equation}
  \alpha_{\mathrm{active,eff}}
  =
  \frac{\alpha_{\mathrm{active}}}
  {1+I/I_{\mathrm{sat}}}.
  \label{eq:active_saturation}
\end{equation}
At low intensity, $I\ll I_{\mathrm{sat}}$, the effective absorption remains close to its unsaturated value. At intensities comparable to or larger than $I_{\mathrm{sat}}$, the absorption grows more slowly with intensity because a significant fraction of the active centers is already driven by the optical field.

This saturation affects only the active spectral channel. It reduces the active absorbed power that could generate laser cooling, but it also reduces active heating when the active transition is thermodynamically unfavorable. Parasitic background absorption is not automatically saturated in this model. Therefore, at high optical intensity, saturation can make parasitic heating relatively more important than the useful active absorption channel.

\subsection{Phonon-assisted absorption}

The simulator includes an optional phenomenological correction for phonon-assisted absorption \cite{pringsheim1929,sheik2009,seletskiy2016}. This correction is intended to represent absorption processes that depend on the thermal occupation of vibrational modes, such as anti-Stokes absorption tails or transitions that require phonon participation. In this mode, the active absorption is multiplied by a temperature-dependent factor,
\begin{equation}
  \alpha_{\mathrm{active}}(\lambda,T)
  \longrightarrow
  \alpha_{\mathrm{active}}(\lambda,T)
  F_{\mathrm{ph}}(\lambda,T).
  \label{eq:phonon_abs_factor}
\end{equation}

The phonon population is modeled using a Bose--Einstein occupation number for a mode with energy $E_{\mathrm{ph}}$ expressed in $\mathrm{cm}^{-1}$,
\begin{equation}
  n_{\mathrm{ph}}(T)
  =
  \frac{1}
  {
  \exp\left[
  \frac{E_{\mathrm{ph}}}{0.69503476\,T}
  \right]-1
  }.
  \label{eq:nph_bose}
\end{equation}
The numerical factor $0.69503476$ is the value of $\kb/(hc)$ in units of $\mathrm{cm}^{-1}\,\mathrm{K}^{-1}$, so that $0.69503476\,T$ is the thermal energy expressed in wavenumber units. The multiplicative correction is normalized to room temperature,
\begin{equation}
  F_{\mathrm{ph}}(\lambda,T)
  =
  \left[
  \frac{n_{\mathrm{ph}}(T)}
  {n_{\mathrm{ph}}(300\,\mathrm{K})}
  \right]^{p(\lambda)}.
  \label{eq:fph}
\end{equation}
With this normalization, the room-temperature absorption is left unchanged, while the absorption can increase or decrease as the particle heats or cools.

The exponent $p(\lambda)$ controls the sensitivity of the active absorption to phonon occupation. If $p=0$, the correction has no effect. If $p>0$, the absorption becomes stronger when the relevant phonon population increases. This model is intentionally phenomenological. It does not assign a complete microscopic multiphonon transition network, but provides a controlled way to test how a temperature-dependent anti-Stokes absorption tail influences the thermal balance.

\subsection{Mean fluorescence wavelength}

The active optical channel depends not only on how much light is absorbed, but also on the mean energy carried away by the emitted fluorescence photons. Since photon energy is $E=hc/\lambda$, the relevant average is an energy-weighted average, not a simple arithmetic average of wavelength. The simulator represents this energy balance through the mean fluorescence wavelength $\lambdaf(T)$ \cite{mccumber1964,sheik2009,seletskiy2016}.

When an emission spectrum $S(\lambda,T)$ is available, the mean fluorescence wavelength is calculated as
\begin{equation}
  \lambdaf(T)
  =
  \frac{\int S(\lambda,T)\,d\lambda}
  {\int S(\lambda,T)\lambda^{-1}\,d\lambda}.
  \label{eq:lambdaf}
\end{equation}
This definition is equivalent to first computing the mean emitted photon energy,
\begin{equation}
  \langle E_{\mathrm{em}}\rangle
  =
  hc\,
  \frac{\int S(\lambda,T)\lambda^{-1}\,d\lambda}
  {\int S(\lambda,T)\,d\lambda},
  \label{eq:mean_emission_energy}
\end{equation}
and then defining $\lambdaf$ through $\langle E_{\mathrm{em}}\rangle=hc/\lambdaf$.

The same emission spectrum that appears in the luminescence plot is used to calculate $\lambdaf(T)$. Therefore, the plotted spectrum is not merely a visualization; it is the spectral input that determines the active optical cooling efficiency. If the user loads a CSV table with luminescence data or changes the generated emission bands, the displayed spectrum and the calculated mean fluorescence wavelength change consistently.

\subsection{Optical cooling efficiency}

The optical cooling efficiency compares the mean energy emitted by fluorescence with the laser photon energy absorbed by the active transition. In the simulator it is modeled as \cite{sheik2009,opticalrefrigerationbook,seletskiy2016}
\begin{equation}
  \etac(\lambda_L,T)
  =
  \etaext
  \frac{\lambda_L}{\lambdaf(T)}
  -1,
  \label{eq:eta_cooling}
\end{equation}
where $\lambda_L$ is the laser wavelength and $\etaext$ is the external quantum efficiency. This form follows from the inverse relation between photon energy and wavelength. If the fluorescence is blue-shifted relative to the laser, $\lambdaf<\lambda_L$, the emitted photons can carry away more energy than was provided by the absorbed laser photons. If the external quantum efficiency is sufficiently high, the excess energy is extracted from the internal degrees of freedom of the particle.

When $\etac>0$, active absorption produces net cooling. When $\etac<0$, active absorption produces net heating. Negative efficiency can occur if the emission is Stokes-shifted, if the external quantum efficiency is too low, or if radiative extraction is not efficient enough to compensate the absorbed optical energy. The active cooling power from all enabled optical tweezers is computed as
\begin{equation}
  \Pcool
  =
  \sum_i
  \left[
  \Pabsact^{(i)}
  \etac^{(i)}
  -
  \Pabsbg^{(i)}
  \right],
  \label{eq:pcool_sum}
\end{equation}
where the index $i$ labels the active tweezers. The first term represents the useful cooling or heating associated with active absorption and fluorescence. The second term subtracts parasitic heating because background absorption always deposits heat in the particle.

The corresponding thermal power entering the internal temperature equation is
\begin{equation}
  \dot Q_{\mathrm{laser}}
  =
  -\Pcool.
  \label{eq:q_laser}
\end{equation}
Thus, $\Pcool>0$ gives $\dot Q_{\mathrm{laser}}<0$ and lowers $\Tint$, while $\Pcool<0$ gives $\dot Q_{\mathrm{laser}}>0$ and heats the particle. This sign convention keeps the optical term consistent with the gas and radiation channels described in the previous sections.

\section{Custom spectra and generated bands}

The simulator separates the mechanical material parameters from the spectral source used in the optical thermal term \cite{mccumber1964,sheik2009,seletskiy2016}. The particle radius, density, refractive index, and charge determine the mass, optical forces, and Paul-trap dynamics. The absorption and emission spectra determine the active optical power, mean fluorescence wavelength, and cooling efficiency. This separation is useful because the built-in presets are generic and are not intended to represent fully calibrated materials. For quantitative modeling of a specific sample, the preferred approach is to load experimentally measured absorption and emission spectra by CSV.

\subsection{CSV spectral tables}

The simulator accepts CSV tables containing a wavelength column and at least one useful spectral column. The recommended template is
\begin{verbatim}
temperature_K,wavelength_nm,alpha_abs_cm_inv,sigma_abs_cm2,
sigma_em_cm2,fluorescence_intensity
\end{verbatim}
The column \texttt{wavelength\_nm} defines the spectral mesh. The optional column \texttt{temperature\_K} allows the user to provide spectra measured at different temperatures. The columns \texttt{alpha\_abs\_cm\_inv} and \texttt{sigma\_abs\_cm2} describe absorption, either as a macroscopic absorption coefficient or as a microscopic cross section. The columns \texttt{sigma\_em\_cm2} and \texttt{fluorescence\_intensity} describe emission or luminescence.

A table containing both absorption and emission allows the simulator to compute both $\alpha_{\mathrm{active}}(\lambda,T)$ and $\lambdaf(T)$ from the same data source. If the table contains absorption but no emission, the active absorbed power can still be computed, but the fluorescence spectrum must be supplied by another model or replaced by a fallback mean fluorescence wavelength. If the table contains emission but no absorption, the emission can still determine $\lambdaf(T)$, while the absorption must be supplied by generated bands or another absorption source.

When multiple temperatures are present in the table, the simulator interpolates between adjacent spectral datasets. This makes it possible to represent temperature-dependent spectral shifts, broadening, or changes in relative intensity. When only one temperature is provided, or when no temperature column is used, the spectrum is treated as an effective temperature-independent input.

For quantitative laser-cooling simulations, absolute absorption data are preferable. A measured absorption coefficient in $\mathrm{cm}^{-1}$ or a calibrated absorption cross section in $\mathrm{cm}^2$ gives a physically meaningful absorbed power. Emission in arbitrary units can still be useful for calculating $\lambdaf(T)$, provided that the relative spectral shape is reliable, because the mean fluorescence wavelength depends on the normalized spectral distribution rather than on the absolute emission amplitude.

\subsection{Generated Gaussian bands}

The generated-band model provides a phenomenological way to construct absorption or emission spectra when measured data are not available. The spectrum is represented as a sum of Gaussian bands \cite{mccumber1964,sheik2009,seletskiy2016},
\begin{equation}
  S(\lambda,T)
  =
  \sum_{i=1}^{N_b}
  A_i(T)
  \exp\left[
  -\frac{(\lambda-\lambda_i)^2}{2\sigma_i(T)^2}
  \right].
  \label{eq:generated_bands}
\end{equation}
Each band is characterized by a center wavelength $\lambda_i$, a width $\sigma_i$, an amplitude $A_i$, and a thermal activation energy $E_i$. The center determines where the band appears in the spectrum, the width determines its spectral spread, and the amplitude determines its relative weight. The thermal activation energy controls how strongly the band amplitude changes with temperature.

The temperature-dependent amplitude is modeled as
\begin{equation}
  A_i(T)
  =
  A_i(300\,\mathrm{K})
  \frac{
  \exp\left[-E_i/(0.69503476\,T)\right]
  }
  {
  \exp\left[-E_i/(0.69503476\cdot300\,\mathrm{K})\right]
  }.
  \label{eq:thermal_amplitude_band}
\end{equation}
If $E_i=0$, the amplitude is temperature independent. If $E_i>0$, the band is thermally activated: it becomes stronger at higher temperature and weaker as the particle cools. This behavior can mimic, in a simplified way, the thermal population of sublevels or vibrationally assisted states.

The width can also vary with temperature according to
\begin{equation}
  \sigma_i(T)
  =
  \sigma_i(300\,\mathrm{K})
  \left[
  1
  +
  b
  \left(
  \sqrt{\frac{T}{300\,\mathrm{K}}}
  -1
  \right)
  \right],
  \label{eq:thermal_width_band}
\end{equation}
with internal numerical limits used to avoid negative or unstable widths. The parameter $b$ controls whether the band broadens or narrows as the temperature changes. This is a phenomenological way to represent the common tendency of solid-state spectra to broaden at higher temperature because of stronger phonon coupling.

The energy $E_i$, expressed in $\mathrm{cm}^{-1}$, does not shift the center of the band. It only controls the thermal amplitude of that spectral component. In an absorption spectrum, increasing $E_i$ makes the absorption band more temperature dependent. In an emission spectrum, increasing $E_i$ makes that emission component less important at low temperature. Thus, $E_i$ should be interpreted as a thermal-activation scale rather than as an automatic assignment to a specific microscopic material level.

Generated bands are useful for exploring how spectral centers, widths, amplitudes, and temperature-dependent populations affect the thermal dynamics. They are not a substitute for calibrated material spectra. When reliable experimental data are available, CSV spectra should be used for quantitative predictions, while generated bands should be viewed as a controlled phenomenological tool for qualitative studies and parameter scans.

\section{Numerical integration}

The time dynamics implemented in the simulator is solved by an explicit fixed-step integrator \cite{higham2001,kloeden1992,hairer2003}. The physical time step is denoted by $\Delta t$ and is chosen by the user in the execution panel. This step controls the actual evolution of the differential equations; visualization parameters, such as decimation of the points displayed in the graphs, do not alter the integrated physical dynamics, only the frequency with which data are stored and drawn.

At each time step, the simulator updates, for each nanoparticle, the set of dynamical variables
\begin{equation}
  \left\{
  x,y,z,\,
  v_x,v_y,v_z,\,
  \Tint,\,
  \Tcom
  \right\}.
\end{equation}
The internal temperature $\Tint$ represents the internal thermal degree of freedom of the nanoparticle, associated with vibrational energy, optical absorption, thermal emission, and heat exchange with the gas. By contrast, $\Tcom$ is an effective kinetic temperature associated with the center-of-mass motion. 

\subsection{General structure of a time step}

The operational order of an integration step is the following. First, interparticle forces are calculated when Coulomb interactions are enabled. Then, for each particle, the optical forces, local optical intensity, internal-temperature heat-exchange terms, manual damping and feedback forces, and finally the mechanical accelerations are calculated. After that, the velocities are updated, the positions are advanced using the new velocities, and the effective center-of-mass kinetic temperature is recalculated.

Schematically, for the step $n\rightarrow n+1$:
\begin{align}
  \mathbf{F}_{\mathrm{pair}}^n
    &\longrightarrow
  \mathbf{F}_{\mathrm{opt}}^n,\ I_{\mathrm{loc}}^n
    \longrightarrow
  \dot{\Tint}^{\,n}
    \longrightarrow
  \mathbf{F}_{\mathrm{fb}}^n,\ \mathbf{F}_{\mathrm{man}}^n
    \longrightarrow
  \mathbf{a}^{\,n}
    \longrightarrow
  \mathbf{v}^{\,n+1}
    \longrightarrow
  \mathbf{r}^{\,n+1}
    \longrightarrow
  \Tcom^{n+1}.
\end{align}
This choice makes the algorithm simple, fast, and suitable for real-time visualization, at the cost of requiring $\Delta t$ to be sufficiently small to resolve micromotion, secular oscillations, and relevant variations of the optical forces.

\subsection{Updating the internal temperature}

The internal temperature is integrated by explicit Euler \cite{hairer2003}. The simulator calculates a total rate
\begin{equation}
  \dot{\Tint}
  =
  \dot{\Tint}_{\mathrm{gas}}
  +
  \dot{\Tint}_{\mathrm{rad}}
  +
  \dot{\Tint}_{\mathrm{laser}},
\end{equation}
where the three terms respectively represent heat exchange with the gas, radiative exchange with the external thermal bath, and light-induced heating or cooling.

When the thermal model is written in terms of powers, the temperature rate can be interpreted as
\begin{equation}
  \dot{\Tint}
  =
  \frac{
  P_{\mathrm{gas}}
  +
  P_{\mathrm{rad}}
  +
  P_{\mathrm{laser}}
  }{C_{\mathrm{th}}},
\end{equation}
where $C_{\mathrm{th}}$ is the effective heat capacity of the particle. For a homogeneous particle,
\begin{equation}
  C_{\mathrm{th}} = m c_p,
\end{equation}
while, in the case of a core--shell geometry, the simulator uses a composite heat capacity,
\begin{equation}
  C_{\mathrm{th}}
  =
  \rho_{\mathrm{core}}V_{\mathrm{core}}c_{p,\mathrm{core}}
  +
  \rho_{\mathrm{shell}}V_{\mathrm{shell}}c_{p,\mathrm{shell}}.
\end{equation}

The explicit update is then
\begin{equation}
  \Tint^{n+1}
  =
  \max\left[
  10^{-9},
  \Tint^n+\Delta t\,\dot{\Tint}^{\,n}
  \right].
\end{equation}
The lower bound has no direct physical meaning; it is a numerical protection against negative or zero temperatures when the time step is too large or when some thermal term produces a very abrupt variation.

The gas term can operate in different regimes. In legacy mode, it is proportional to the difference $\Tint-T_{\mathrm{gas}}$. In physical mode, exchange can be treated by conduction in the intermediate regime or by an expression for the free-molecular regime, depending on the Knudsen number. The radiative term can be disabled, treated by gray-body Stefan--Boltzmann, or calculated by spectral Rayleigh/Mie models. The optical term can represent both heating by parasitic absorption and anti-Stokes cooling, depending on the material, absorption/emission spectrum, external quantum efficiency, and local intensity.

\subsection{Effective thermal bath of the center of mass}

For the numerical integration of the translational degrees of freedom, the stochastic velocity diffusion is not computed from the diagnostic temperature $\Tcom$ itself. Instead, the integrator uses the temperature of the effective bath coupled to the center-of-mass motion. This bath combines the gas environment and, when enabled, the phenomenological internal--COM coupling introduced in the previous sections.

The total physical damping entering the Langevin step is
\begin{equation}
  \Gamma_{\mathrm{phys}}
  =
  \Gamma_{\mathrm{gas}}
  +
  \Gamma_{\mathrm{int}},
\end{equation}
where
\begin{equation}
  \Gamma_{\mathrm{int}}
  =
  \chi_{\mathrm{int}\rightarrow\mathrm{COM}}
  \Gamma_{\mathrm{gas}}.
\end{equation}
The corresponding effective bath temperature is
\begin{equation}
  T_{\mathrm{bath}}^{\mathrm{COM}}
  =
  \frac{
  \Gamma_{\mathrm{gas}}\Tgas
  +
  \Gamma_{\mathrm{int}}\Tint
  }{
  \Gamma_{\mathrm{phys}}
  },
  \qquad
  \Gamma_{\mathrm{phys}}>0.
  \label{eq:tcom_effective_bath}
\end{equation}
Thus, the stochastic force samples a weighted thermal reservoir: the gas contributes with weight $\Gamma_{\mathrm{gas}}$, while the internal temperature contributes with weight $\Gamma_{\mathrm{int}}$. If the internal--COM coupling is disabled, Eq.~\eqref{eq:tcom_effective_bath} reduces to $T_{\mathrm{bath}}^{\mathrm{COM}}=\Tgas$.

In the velocity update, this effective bath appears through the Langevin diffusion amplitude,
\begin{equation}
  \sigma_v
  =
  \sqrt{
  \frac{
  2\kb
  \Gamma_{\mathrm{phys}}
  T_{\mathrm{bath}}^{\mathrm{COM}}
  }{m}
  }.
  \label{eq:velocity_noise_effective_bath}
\end{equation}
Equivalently,
\begin{equation}
  \sigma_v
  =
  \sqrt{
  \frac{
  2\kb
  \left(
  \Gamma_{\mathrm{gas}}\Tgas
  +
  \Gamma_{\mathrm{int}}\Tint
  \right)
  }{m}
  }.
  \label{eq:velocity_noise_expanded}
\end{equation}
This form ensures consistency between viscous damping and stochastic diffusion in the translational dynamics. The role of $\Tcom$ is therefore diagnostic: it is estimated from the simulated kinetic energy, while the random kicks are determined by the thermal reservoirs coupled to the center-of-mass motion.

\subsection{Updating velocities}

For each Cartesian axis, the velocity is updated by an Euler--Maruyama form \cite{higham2001,kloeden1992}, given by
\begin{equation}
  v_j^{n+1}
  =
  v_j^n
  +
  a_j^n\Delta t
  +
  \sigma_v^n\sqrt{\Delta t}\,\xi_j^n,
  \qquad
  j\in\{x,y,z\},
\end{equation}
where $\xi_j^n$ are independent standard normal random variables, with zero mean and unit variance
\begin{equation}
  \langle \xi_j^n\rangle=0,
  \qquad
  \langle \xi_i^n\xi_j^m\rangle=\delta_{ij}\delta_{nm}.
\end{equation}

The total acceleration contains a deterministic part from the Paul trap, an optical part, an interparticle-interaction part, physical damping, manual damping, and feedback and can be written as
\begin{equation}
  a_j
  =
  a_j^{\mathrm{Paul}}
  -
  \Gamma_{\mathrm{phys}}v_j
  +
  \frac{
  F_j^{\mathrm{opt}}
  +
  F_j^{\mathrm{pair}}
  +
  F_j^{\mathrm{man}}
  +
  F_j^{\mathrm{fb}}
  }{m}.
\end{equation}

The contribution of the Paul trap is applied only while the particle remains inside the geometric region of the trap. Introducing the indicator function
\begin{equation}
  \Theta_{\mathrm{trap}}
  =
  \begin{cases}
  1, & \sqrt{x^2+y^2}\leq r_0\ \mathrm{and}\ |z|\leq z_0,\\
  0, & \text{otherwise},
  \end{cases}
\end{equation}
the deterministic accelerations of the trap can be written, in the form used by the simulator, as
\begin{align}
  a_x^{\mathrm{Paul}}
  &=
  \Theta_{\mathrm{trap}}
  \left[
  -\left(A\cos\Omega t-B\right)x
  +
  a_{\mathrm{comp},x}
  \right],
  \\
  a_y^{\mathrm{Paul}}
  &=
  \Theta_{\mathrm{trap}}
  \left[
  -\left(-A\cos\Omega t-B\right)y
  +
  a_{\mathrm{comp},y}
  \right],
  \\
  a_z^{\mathrm{Paul}}
  &=
  \Theta_{\mathrm{trap}}
  \left[
  -\omega_z^2 z
  +
  a_{\mathrm{ec},z}
  \right].
\end{align}
In these expressions, $A$ contains the radial quadrupole AC force, $B$ contains the DC curvature associated with the endcaps, $\Omega$ is the angular frequency of the AC voltage, $a_{\mathrm{comp},x}$ and $a_{\mathrm{comp},y}$ are accelerations produced by radial DC compensation fields, and $a_{\mathrm{ec},z}$ represents the axial displacement produced by endcap imbalance.

The optical force $\mathbf{F}^{\mathrm{opt}}$ is calculated from the local intensity of each optical tweezer. The simulator sums the contributions of all enabled beams
\begin{equation}
  \mathbf{F}^{\mathrm{opt}}
  =
  \sum_{\ell}
  \mathbf{F}_{\ell}^{\mathrm{opt}}.
\end{equation}
Each beam contributes terms associated with the transverse/longitudinal intensity gradient and the scattering force along the beam propagation direction.

The pair force $\mathbf{F}^{\mathrm{pair}}$ represents the Coulomb interaction between charged nanoparticles. For particle $i$, it can be written schematically as
\begin{equation}
  \mathbf{F}^{\mathrm{pair}}_i
  =
  \sum_{k\neq i}
  \frac{1}{4\pi\epsilon_0}
  q_iq_k
  \frac{\mathbf{r}_i-\mathbf{r}_k}
  {\left(|\mathbf{r}_i-\mathbf{r}_k|^2+\epsilon_{\mathrm{soft}}^2\right)^{3/2}},
\end{equation}
with possible short-range softening by $\epsilon_{\mathrm{soft}}$ and maximum-force limiting to avoid numerical instabilities at very small separations.

Manual damping is implemented as a deterministic viscous force,
\begin{equation}
  F_j^{\mathrm{man}}
  =
  -m\Gamma_{\mathrm{man},j}v_j.
\end{equation}
This term does not add thermal noise. Therefore, it should be interpreted as a dissipative actuator imposed by the user, not as a physical thermal bath automatically satisfying a fluctuation--dissipation relation.

Automatic feedback also acts as cold damping, 
\begin{equation}
  F_j^{\mathrm{fb}}
  =
  -m\Gamma_{\mathrm{fb}}v_j,\label{feedback_force}
\end{equation}
but only in enabled axes and only when the conditions imposed by the user are satisfied. The simulator allows the maximum feedback force to be limited, the actuation to be stopped below a velocity threshold, and feedback to be restricted to an ellipsoidal region of the trap. When the spatial gate is active, feedback acts only if
\begin{equation}
  \frac{x^2+y^2}{(\eta_r r_0)^2}
  +
  \frac{z^2}{(\eta_z z_0)^2}
  \leq 1.
\end{equation}

\subsection{Updating positions}

After the velocities are updated, the positions are advanced using the newly calculated velocities
\begin{equation}
  x_j^{n+1}
  =
  x_j^n
  +
  v_j^{n+1}\Delta t,
  \qquad
  j\in\{x,y,z\}.
\end{equation}
This choice corresponds to a semi-implicit Euler scheme, or symplectic Euler, for the mechanical part. The force and acceleration are evaluated at the old state, but the position is updated with the new velocity. For oscillatory systems, this procedure is usually more numerically robust than fully explicit Euler, especially when there are restoring forces such as those of the Paul trap and optical tweezer.

\subsection{Updating \texorpdfstring{$\Tcom$}{TCOM}}

After each velocity update, the simulator computes the instantaneous kinetic temperature associated with the center-of-mass motion,
\begin{equation}
  T_{\mathrm{kin}}^{n+1}
  =
  \frac{
  m\left[
  \left(v_x^{n+1}\right)^2+
  \left(v_y^{n+1}\right)^2+
  \left(v_z^{n+1}\right)^2
  \right]
  }{
  3\kb
  }.
  \label{eq:tkin_discrete}
\end{equation}
This quantity is evaluated directly from the simulated velocities and is therefore a diagnostic of the instantaneous translational kinetic energy. It is not used as the temperature of the stochastic bath in the next Langevin step.

Because $T_{\mathrm{kin}}$ contains rapid contributions from micromotion, secular oscillations, and stochastic kicks, the displayed center-of-mass temperature is updated as a low-pass filtered estimator. The simulator uses the discrete update
\begin{equation}
  \Tcom^{n+1}
  =
  \max
  \left[
  T_{\mathrm{floor}},
  \left(1-\alpha_{\mathrm{COM}}\right)\Tcom^n
  +
  \alpha_{\mathrm{COM}}T_{\mathrm{kin}}^{n+1}
  \right],
  \label{eq:tcom_filter_update}
\end{equation}
where
\begin{equation}
  \alpha_{\mathrm{COM}}
  =
  1-\exp\left(-\gamma_{\mathrm{smooth}}\Delta t\right).
  \label{eq:tcom_filter_alpha}
\end{equation}
The exponential form keeps the filter numerically stable for any positive value of $\gamma_{\mathrm{smooth}}\Delta t$, while recovering the Euler form
\begin{equation}
  \alpha_{\mathrm{COM}}
  \simeq
  \gamma_{\mathrm{smooth}}\Delta t
\end{equation}
when $\gamma_{\mathrm{smooth}}\Delta t\ll1$.

The smoothing rate is chosen as a positive diagnostic rate rather than as a purely gas-limited relaxation rate,
\begin{equation}
  \gamma_{\mathrm{smooth}}
  =
  \max
  \left[
  \gamma_{\mathrm{diag}},
  2\Gamma_{\mathrm{COM}}^{\mathrm{appl}},
  10^{-6}\,\mathrm{s}^{-1}
  \right].
  \label{eq:gamma_smooth_numeric}
\end{equation}
Here $\Gamma_{\mathrm{COM}}^{\mathrm{appl}}$ denotes the effective damping rate actually applied to the center-of-mass velocity update, including physical gas damping and any enabled deterministic damping channel, such as feedback or manual cooling. The term $\gamma_{\mathrm{diag}}$ is a diagnostic floor used to prevent the displayed $\Tcom$ from becoming artificially frozen when the gas pressure is extremely low.

The role of $\Tcom$ in the integrator is therefore limited to monitoring and visualization. The stochastic diffusion amplitude is computed from the effective bath coupled to the translational motion, whereas $\Tcom$ is reconstructed from the simulated velocities after the update. This separation avoids a circular numerical definition in which the diagnostic temperature would both measure and impose the center-of-mass fluctuations.
\subsection{Sampling, visualization, and numerical stability}

At each step, the simulated time is incremented by
\begin{equation}
  t^{n+1}=t^n+\Delta t.
\end{equation}
However, not all steps need to be sent to the graphs. The simulator can save and display only one point every certain number of steps, reducing the graphical cost without altering the physical integration.

The stability of the method depends on the choice of $\Delta t$. The step must be small enough to resolve the AC frequency $\Omega$, the time scale of the secular oscillations, the variation scale of the optical forces, and the characteristic damping times. If $\Delta t$ is too large, numerical heating, artificial growth of the micromotion amplitude, errors in the estimate of $\Tcom$, or instabilities in the internal temperature may appear. Therefore, quantitative interpretation of the results should always check convergence under reduction of $\Delta t$, especially in regimes of high AC voltage, strong optical tweezer, high charge, strong feedback, or Coulomb interaction between multiple particles.

\section{Model limitations}

The simulator was built to combine, in real time, the dynamics of a charged nanoparticle in a Paul trap, the action of optical tweezers, the internal-temperature balance, heat exchange with the gas, radiative emission, and spectral laser-cooling models. This breadth makes the model useful for physical exploration and for visualizing regimes, but it also imposes limitations. The main goal is not to replace a complete microscopic simulation of all solid-state degrees of freedom, but to provide a reduced description that keeps the main channels of force, dissipation, noise, and thermal balance.

The first important limitation is that the spectra internally generated by the simulator are phenomenological. They reproduce qualitative features of absorption and emission bands, such as center position, width, amplitude, thermal broadening, and activation by effective energy, but they do not automatically represent a specific material. Even the presets should be read as references for exploring plausible scenarios, not as universal databases. For a prediction to be quantitative, it is necessary to use calibrated experimental spectra, preferably containing absorption and emission in the same physical scale relevant for the material, dopant concentration, polarization, temperature, and sample geometry.

The internal temperature $\Tint$ is also an effective variable. The simulator does not resolve spatial temperature profiles inside the particle, nor does it calculate gradients between core, shell, surface, and external medium. Even in core--shell mode, there is a single internal temperature common to the system. The core--shell separation alters active volume, heat capacity, mass, and phenomenological surface factors, but it does not introduce two independent temperatures. Therefore, situations in which the core and shell are strongly out of thermal equilibrium, or where internal gradients are relevant, are not explicitly described.

The coupling between internal temperature and center of mass is another deliberately phenomenological approximation. The parameter $\chi_{\mathrm{int\to COM}}$ makes it possible to study how an internally hot or cold particle could modify the effective mechanical bath of translational motion, but it does not derive from a complete microscopic theory. Quantitative interpretation of this coupling would require an additional model for the real mechanisms of conversion between internal and translational energy.

The description of optical forces also has a domain of validity. The tweezer force is treated by a Rayleigh-type polarizable response, appropriate when the particle radius is sufficiently small compared with the wavelength and when the response can be approximated by an effective polarizability. In this approximation, the optical force is decomposed into gradient and scattering contributions, but it does not resolve the complete field distribution inside and around larger particles. For particles whose size is comparable to the wavelength, Mie resonances, shape effects, anisotropy, multiple internal modes, and high-numerical-aperture vector corrections can significantly alter the real force. The simulator includes Rayleigh/Mie models for spectral thermal radiation, but this does not mean that the optical tweezer force is being calculated by a complete Mie solution.

The Coulomb interaction between multiple particles is also treated in a reduced form \cite{paul1990,dani2021}. The model calculates pair forces between effective charges, with softening and limiting to avoid numerical divergences at very short distances. This formulation captures the dominant electrostatic repulsion or attraction between charged particles, but it does not fully include space-charge effects in the trap field. In regimes with many particles, high charges, or very short distances, the interpretation should be qualitative.

The Paul trap is described by effective quadrupole fields and by a geometric actuation region \cite{paul1990}. This choice makes it possible to reproduce micromotion, secular confinement, parametric instability, and dependence on AC voltage, AC frequency, charge, and mass. Real fields, however, may include geometric imperfections, misalignments, higher multipole terms, electronic noise, nonideal phases between electrodes, and couplings between axes. The simulator includes compensation fields and endcap imbalance, but it does not replace a complete three-dimensional electrostatic solution obtained, for example, by finite elements for a specific experimental geometry.

The gas thermal-exchange model uses approximations for the free-molecular regime, intermediate regime, or a legacy form proportional to gas damping \cite{epstein1924,hebestreit2018}. These descriptions are useful for capturing trends with pressure, size, and thermal accommodation, but they depend on effective parameters such as accommodation coefficient, conductivity, and mean free path. Under real experimental conditions, gas composition, surface roughness, adsorbates, contamination, local pressure, and background heating can alter the thermal exchange. For this reason, thermal accommodation should be viewed as an effective physical parameter, not as a universal constant of the particle.

Thermal radiation has a similar limitation \cite{bohren1983,hebestreit2018}. The gray-body model uses a constant effective emissivity, while the spectral Rayleigh/Mie models use approximations for the infrared absorption of the material. In the absence of calibrated complex optical constants in the infrared, the calculated radiative exchange should be interpreted as an order-of-magnitude estimate. 

Spectral laser cooling is sensitive to parameters that, in real experiments, depend on material quality \cite{epstein1995,sheik2009,seletskiy2010,seletskiy2016,xia2021}. The external efficiency $\eta_{\mathrm{ext}}$, parasitic absorption $\alpha_{\mathrm{bg}}$, active-center concentration, saturation, fluorescence reabsorption, and mean emission are treated compactly. This compact treatment is necessary to keep the simulator usable in real time, but it does not resolve all microscopic processes of nonradiative relaxation, energy transfer between dopants, surface quenching, excitation migration, critical concentration, upconversion, or radiative trapping. Thus, cooling results should be evaluated carefully, especially when they depend on small differences between cooling power and parasitic power.

The numerical integration is explicit and fixed-step \cite{higham2001,kloeden1992,hairer2003}. This type of integrator is simple, fast, and suitable for an interactive interface, but it is not unconditionally stable. Large time steps can generate numerical heating, artificial growth of micromotion, errors in the kinetic-temperature estimate, instabilities in strong Coulomb forces, or unphysical variations of $\Tint$. Regimes with high AC voltage, high AC frequency, highly charged particles, intense optical tweezers, strong feedback, or fast optical heating/cooling require smaller steps. For quantitative interpretation, it is necessary to check whether the result converges when $\Delta t$ is reduced.


\section{Literature benchmarks}
The simulator supports data export in two formats: CSV and binary. The CSV format allows users to export either all recorded variables or a reduced set of essential parameters. Although convenient the CSV becomes inefficient for long recordings at high sampling rates, such as those required to resolve fast oscillator frequency dynamics. Under these conditions, the dataset may grow to several gigabytes, and its accumulation in RAM can slow down the simulator or exhaust the available memory.

For large datasets, the binary format provides a more efficient alternative. Each export consists of a binary data file accompanied by a JSON metadata file containing the simulation parameters and information required to interpret the stored variables. Together, these files allow the complete dataset to be reconstructed and analyzed outside the simulator. In this section, exported data are used to reproduce selected results from the literature, benchmark the simulator, and demonstrate its ability to capture trends observed in realistic experimental scenarios.

\subsection{Laser-induced internal heating}
\label{sec:validation_zhang_fig4}

The thermal module was benchmarked against the experiment of Zhang \textit{et al.}~\cite{zhang2023temperature}. In that work, an optically levitated $\alpha$-$\mathrm{NaYF_4:Yb^{3+}/Er^{3+}}$ nanoparticle was trapped with a $1064~\mathrm{nm}$ optical tweezer and simultaneously excited by a $976~\mathrm{nm}$ laser. Its internal temperature was inferred from the luminescence-intensity ratio of the thermally coupled ${}^{2}\mathrm{H}_{11/2}$ and ${}^{4}\mathrm{S}_{3/2}$ levels of $\mathrm{Er^{3+}}$. At $20~\mathrm{mbar}$, the lower excitation powers produced a rapid temperature increase followed by an approximately stationary regime, whereas the highest power caused sustained heating before the particle escaped from the optical trap.

To reproduce this scenario, the simulator calculated the time evolution of the nanoparticle's internal temperature while the power of the $976~\mathrm{nm}$ beam was varied. The dynamics follows the energy-balance equation in Eq.~\eqref{eq:energy_balance_internal}, with the steady or transient response determined by the competition between optical heating and heat dissipation.

\begin{figure}[htbp]
    \centering
    \includegraphics[width=0.6\linewidth]{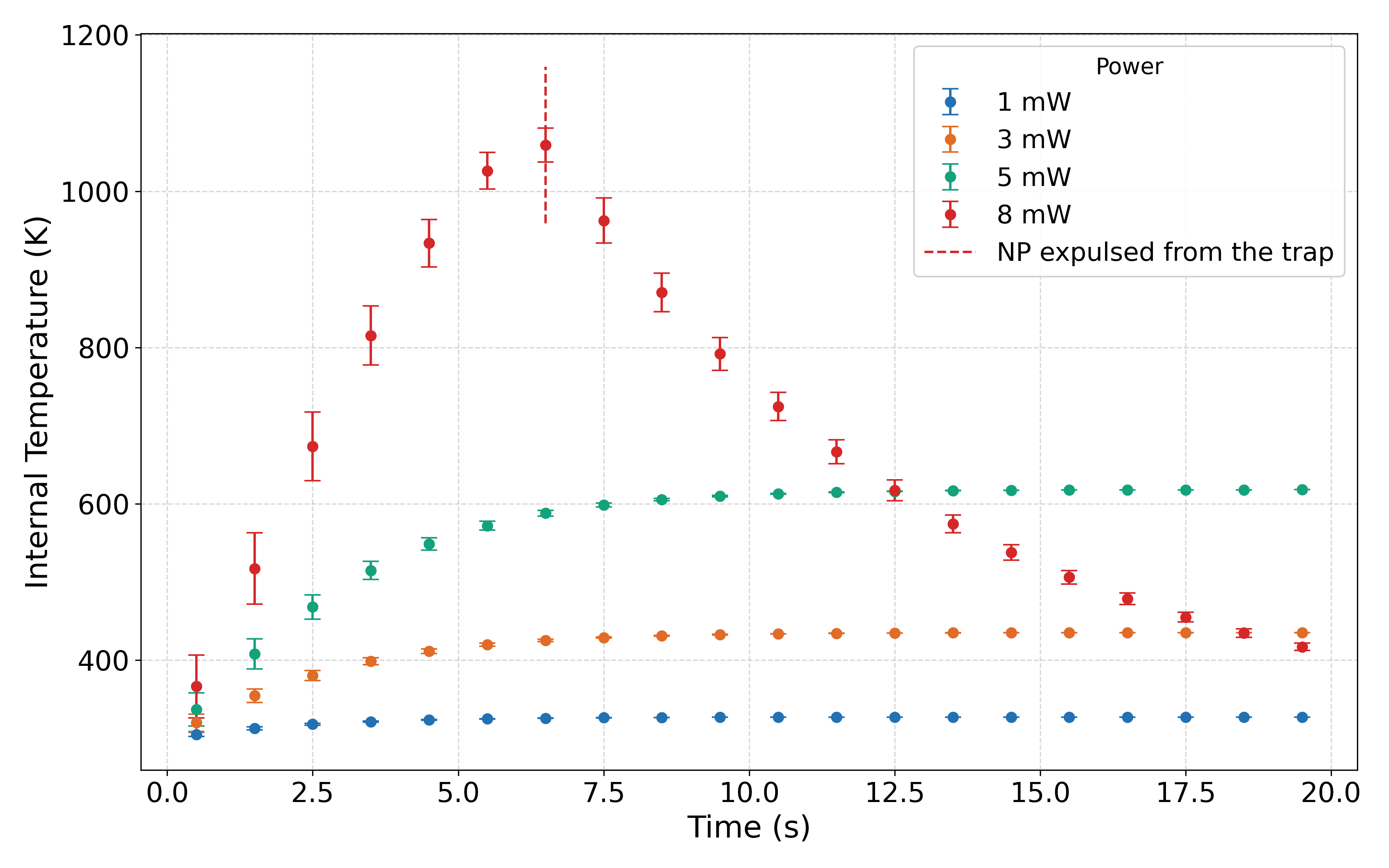}
    \caption{Simulated internal temperature as a function of time for different powers of the $976~\mathrm{nm}$ excitation laser. Lower powers approach quasi-steady temperatures, whereas $8~\mathrm{mW}$ produces a transient high-temperature regime followed by post-escape cooling.}
    \label{fig:simulated_internal_temperature_power}
\end{figure}

Figure~\ref{fig:simulated_internal_temperature_power} shows the simulated response for excitation powers of $1$, $3$, $5$, and $8~\mathrm{mW}$. At $1$, $3$, and $5~\mathrm{mW}$, the temperature rises from its initial value and approaches approximately $325$, $435$, and $618~\mathrm{K}$, respectively. The increase in the quasi-steady temperature with optical power reproduces the main lower-power trend reported in Ref.~\cite{zhang2023temperature}.

At $8~\mathrm{mW}$, the simulated internal temperature reaches approximately $1.1\times10^{3}~\mathrm{K}$. It begins to decrease after the particle leaves the optical-tweezer region, where the local laser intensity and the associated optical heating become negligible. The nanoparticle then relaxes toward the gas temperature. In the experiment, the corresponding instability led to particle loss when $T_{\mathrm{int}}$ exceeded approximately $500~\mathrm{K}$, thereby terminating the measurement. Because the simulator continues to integrate both the trajectory and internal temperature after escape, the descending branch represents post-loss thermal relaxation rather than a trapped-particle state. Thus, the model captures the power-dependent transition from quasi-steady heating to a strongly transient regime followed by trap escape.

\subsection{Coupling between internal and center-of-mass temperatures}
\label{subsec:tint_tcom_coupling}

The internal temperature, $T_{\mathrm{int}}$, and the center-of-mass temperature, $T_{\mathrm{COM}}$, describe distinct degrees of freedom. The former characterizes the internal energy of the nanoparticle, primarily that stored in lattice vibrations, whereas the latter describes its translational motion. They therefore need not be equal, especially under vacuum, where optical absorption can change $T_{\mathrm{int}}$ while feedback acts directly on the center-of-mass dynamics.

Despite this distinction, the two temperatures may be indirectly related through interactions with the residual gas. Molecules leaving an internally heated particle can carry energy acquired at its surface, thereby modifying the fluctuating forces acting on the translational motion. This nonequilibrium mechanism has been described using two-bath models for levitated nanoparticles \cite{millen2014,hebestreit2018}. A correlation between internal and center-of-mass temperatures has also been observed in optically levitated $\mathrm{NaYF_4:Yb^{3+}/Er^{3+}}$ nanoparticles \cite{zhang2023temperature}.

As detailed in the preceding model sections, the simulator represents this effect phenomenologically by coupling the center-of-mass motion to an additional mechanical reservoir characterized by $T_{\mathrm{int}}$. Its strength is controlled by the dimensionless parameter $\chi_{\mathrm{int}\rightarrow\mathrm{COM}}$. When this parameter is nonzero, variations in $T_{\mathrm{int}}$ modify the stochastic excitation of the translational motion and can therefore change $T_{\mathrm{COM}}$.

\begin{figure}[htbp]
  \centering
  \includegraphics[width=0.78\linewidth]{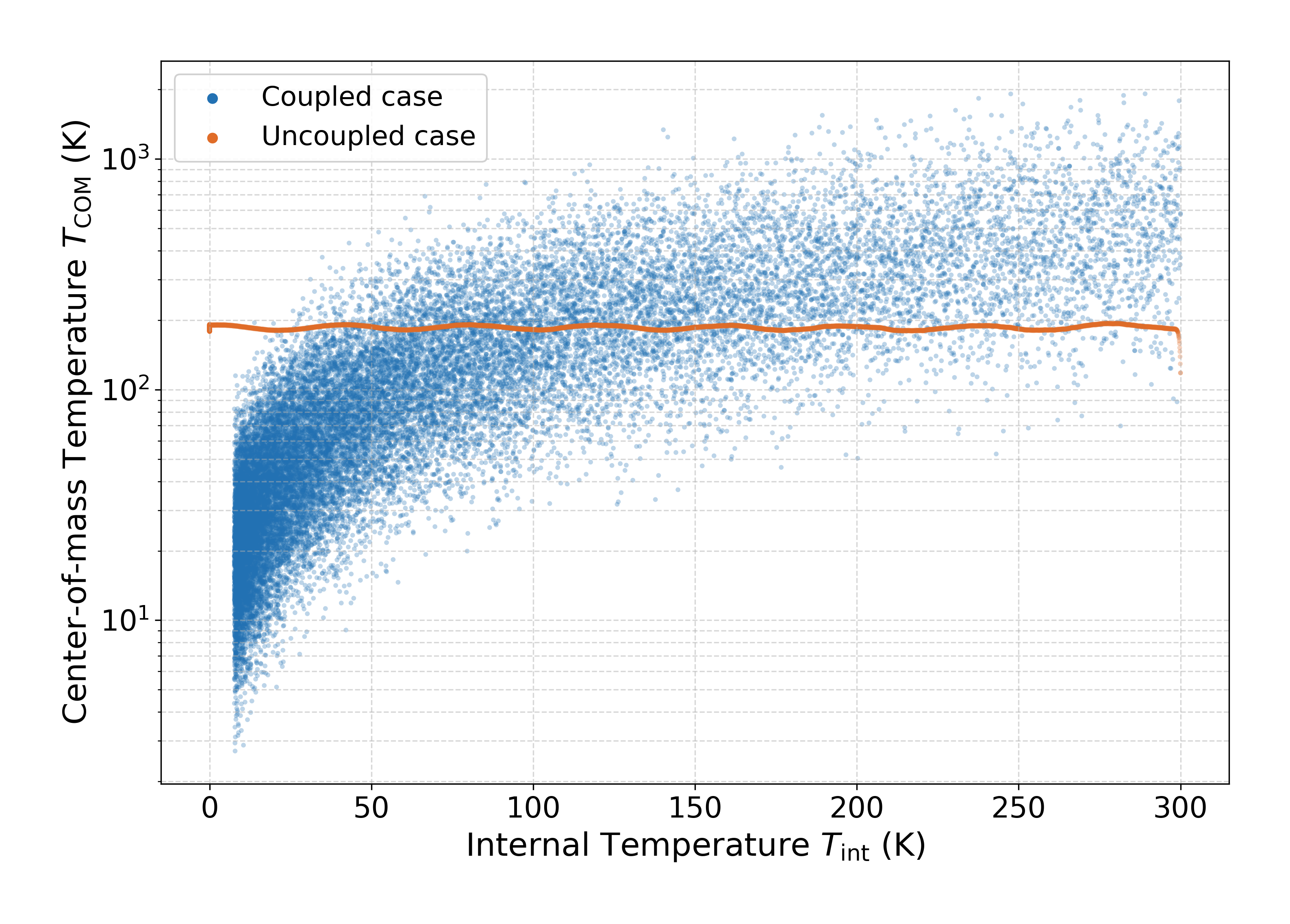}
  \caption{Simulated center-of-mass temperature $T_{\mathrm{COM}}$ as a   function of the internal temperature $T_{\mathrm{int}}$, with and without   internal-to-COM coupling. When the coupling is enabled (blue),   $T_{\mathrm{COM}}$ increases with $T_{\mathrm{int}}$ and exhibits stronger   stochastic fluctuations. When it is disabled (orange),   $T_{\mathrm{COM}}$ remains approximately independent of   $T_{\mathrm{int}}$. The vertical axis is shown on a logarithmic scale.}
  \label{fig:tcom_vs_tint_coupling}
\end{figure}

Figure~\ref{fig:tcom_vs_tint_coupling} shows the resulting numerical behavior. With the coupling enabled, the blue data exhibit a clear positive dependence of $T_{\mathrm{COM}}$ on $T_{\mathrm{int}}$. Their broad distribution reflects the stochastic nature of the motion and the use of an instantaneous kinetic temperature; individual excursions should therefore not be interpreted as stationary temperatures. When the coupling is disabled, the orange trace remains approximately horizontal, confirming that $T_{\mathrm{int}}$ does not directly affect the center-of-mass dynamics in this case.

The implemented coupling is intentionally phenomenological and effectively unidirectional: $T_{\mathrm{int}}$ influences the center-of-mass dynamics, but the corresponding mechanical energy exchange is not returned to the internal energy balance. Moreover, $\chi_{\mathrm{int}\rightarrow\mathrm{COM}}$ is an effective model parameter, not a microscopic gas--surface accommodation coefficient. The comparison in Fig.~\ref{fig:tcom_vs_tint_coupling} should therefore be interpreted as a qualitative demonstration of the coupling implemented in the simulator; quantitative predictions require calibration against experimental data.

\subsection{Collective equilibrium and axial localization of multiple nanoparticles}
\label{sec:collective_phase_space}

The multi-particle capability of the simulator was evaluated using five identical charged nanoparticles confined in a linear Paul trap. Each particle had a radius of $80~\mathrm{nm}$ and a charge of $300e$. In addition to the trapping forces, the model included pairwise Coulomb repulsion, residual-gas damping, stochastic gas collisions, and feedback cooling. Feedback cooling of charged nanoparticles in linear Paul traps has been experimentally demonstrated for both individual particles and Coulomb-coupled systems \cite{dania2021feedback,penny2023sympathetic}. The data analyzed below were recorded only after the particle chain had reached a stationary configuration. Consequently, this benchmark characterizes the resulting collective equilibrium and particle localization rather than the initial loading and chain-formation dynamics.

\begin{figure}[htbp]
    \centering
    \includegraphics[width=0.5\linewidth]{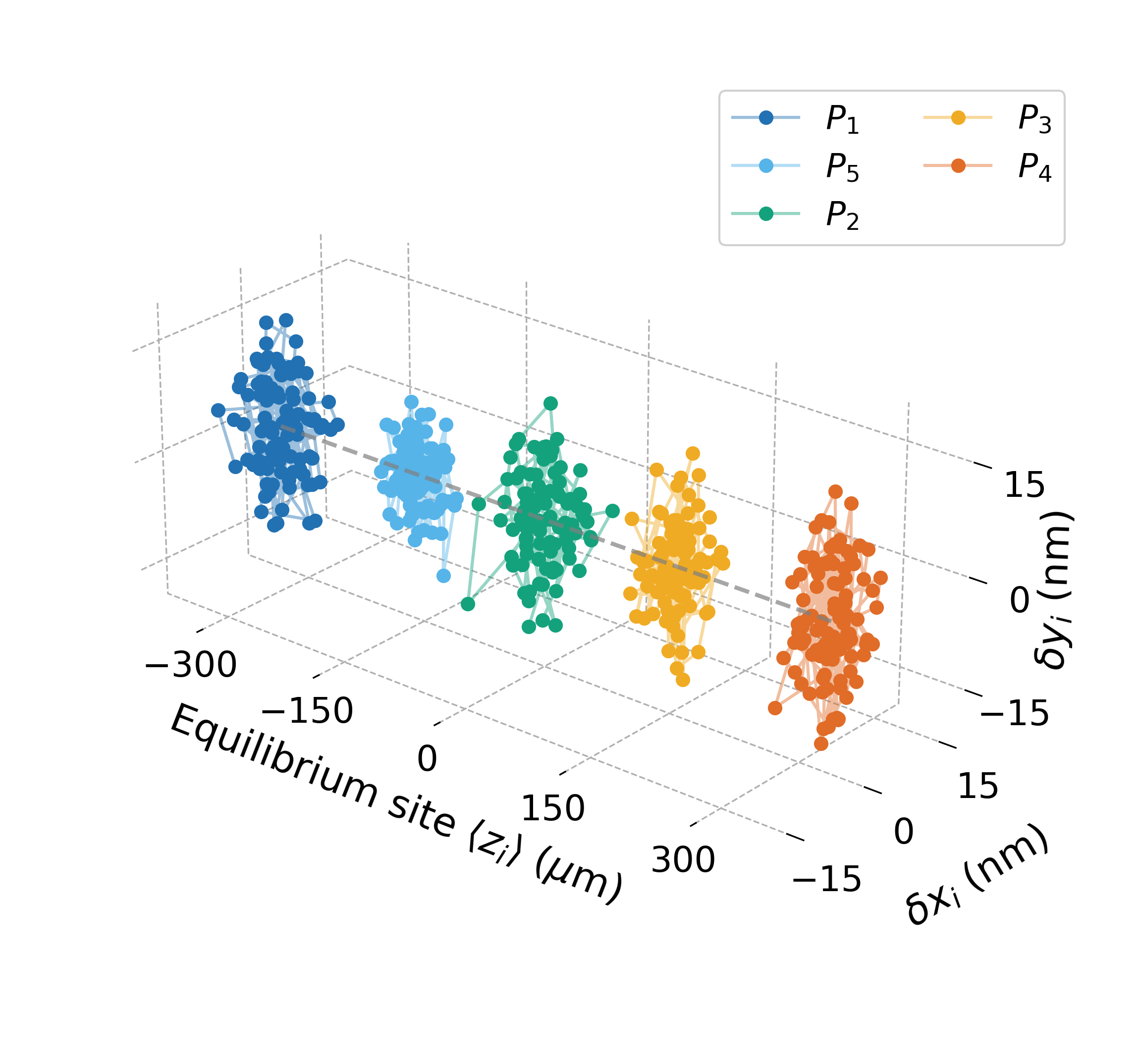}
    \caption{Time-sampled axial position distributions of five identical charged nanoparticles confined in the linear Paul trap. Each narrow peak represents the fluctuations of one particle around its mean equilibrium position after the axial chain has formed. The nonuniform spacing, with the smallest separations near the trap center, results from the self-consistent balance between axial confinement and mutual Coulomb repulsion.}
    \label{fig:collective_particle}
\end{figure}

Figure~\ref{fig:collective_particle} displays five well-resolved distributions centered approximately at
\[
\langle z_i\rangle \simeq
\{-332,\,-156.6,\,0,\,156.6,\,332\}~\mu\mathrm{m}.
\]
The mirror symmetry about $z=0$ indicates the formation of an ordered axial chain centered on the minimum of the trapping potential. The corresponding nearest-neighbor separations are approximately $175.4$, $156.6$, $156.6$, and $175.4~\mu\mathrm{m}$, giving a minimum separation of approximately $156.6~\mu\mathrm{m}$ between the central particles.

This nonuniform spacing is characteristic of a finite Coulomb chain under approximately harmonic axial confinement \cite{james1998}. The particle density is greater near the trap center, whereas the outer particles occupy regions where a stronger axial restoring force is required to balance the accumulated outward Coulomb repulsion. The equilibrium positions are therefore not assigned independently; they emerge self-consistently from the simultaneous force balance involving all five particles.

Gas damping and feedback cooling primarily determine how rapidly the system approaches equilibrium and how broadly each particle fluctuates around its mean position. For ideal velocity-proportional feedback, neither mechanism produces a static force, so the mean configuration remains governed by the trapping potential and Coulomb interaction. This distinction separates the spatial organization of the chain from its motional temperature.

The widths of the distributions are on the order of a few tens of nanometres, more than three orders of magnitude smaller than the minimum interparticle separation. The absence of overlap between neighboring distributions shows that each particle remains strongly localized around a distinct collective equilibrium position. Because the Coulomb force acting on each particle depends on the instantaneous positions of all the others, perturbations of the chain are intrinsically coupled and may be decomposed into collective normal modes. Coulomb-mediated coupling, sympathetic cooling, and the simultaneous control of multiple levitated particles have been investigated experimentally in Paul traps \cite{penny2023sympathetic,ren2025arrays}.

It is important, however, to distinguish collective equilibrium from collective-mode dynamics. The position distributions in Fig.~\ref{fig:collective_particle} demonstrate Coulomb-induced spatial organization and stable localization, but they do not by themselves determine the normal-mode frequencies or quantify dynamical correlations between particles. Such information would require, for example, cross-correlation functions, cross-spectral densities, or a normal-mode decomposition of the individual trajectories. Within its present scope, this result validates the simulator's ability to integrate the coupled motion of several charged particles and recover the expected stationary structure of an ordered Coulomb chain.

\subsection{Feedback cooling}
\label{subsec:feedback_cooling}

The simulator was also used to examine feedback cooling of the nanoparticle's center-of-mass (COM) motion. In the cold-damping scheme, a force approximately opposite to the particle velocity is applied along each controlled direction, as defined in Eq.~\eqref{feedback_force}. This force removes mechanical energy from the corresponding motional mode and lowers its effective COM temperature without changing the background-gas temperature.

\begin{figure}[htbp]
    \centering
    \includegraphics[width=0.72\linewidth]{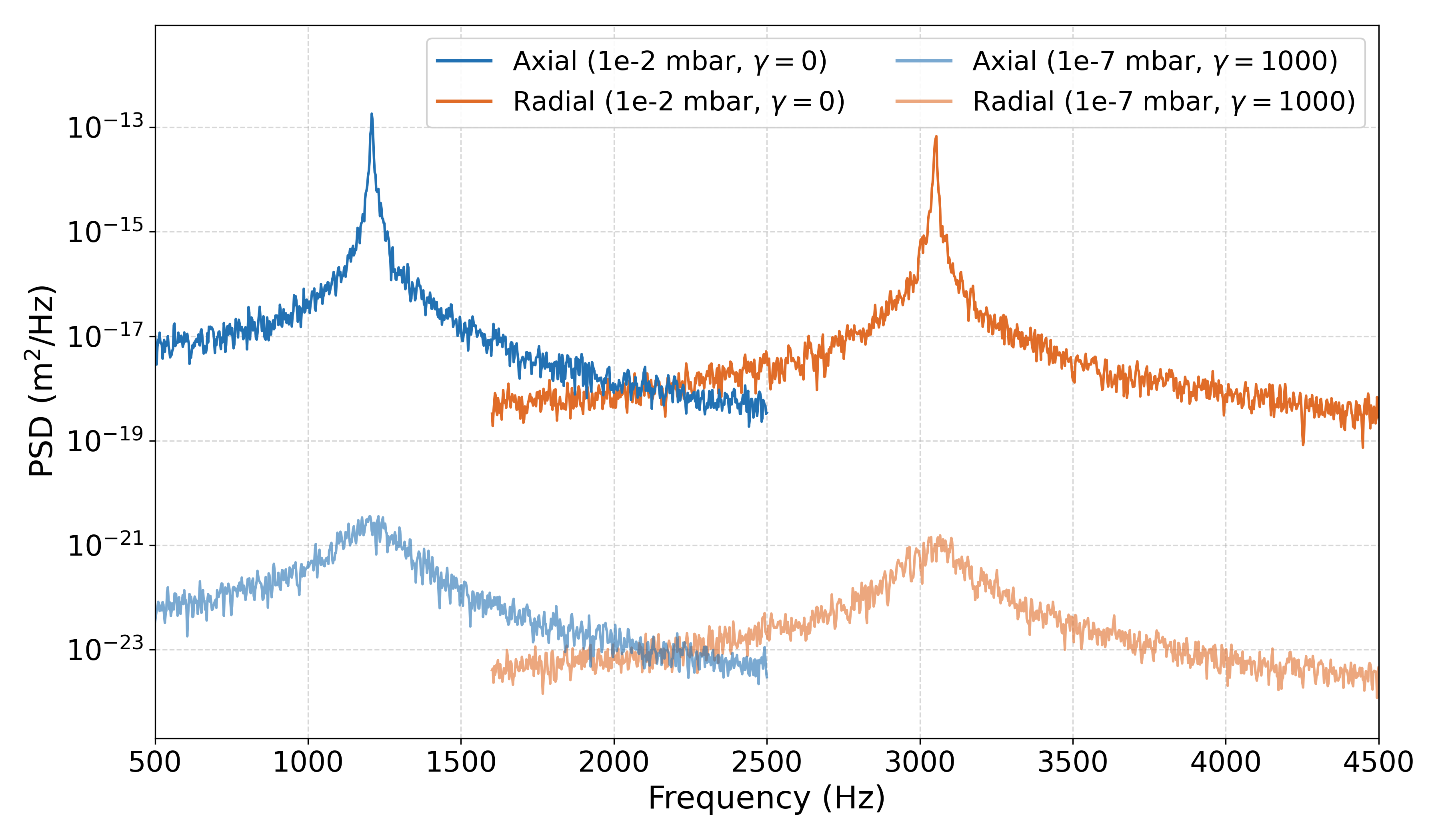}
    \caption{Power spectral density of the nanoparticle's COM motion with feedback off at $10^{-2}~\mathrm{mbar}$ and feedback on at $10^{-7}~\mathrm{mbar}$. In the cooled regime, the motional resonance and its integrated spectral power are strongly suppressed.}
    \label{fig:psd}
\end{figure}

Dania \textit{et al.}~\cite{dani2021} experimentally investigated electrical and optical feedback cooling of a silica nanoparticle in a Paul trap as functions of feedback gain, pressure, and detection conditions. Figure~\ref{fig:psd} compares two representative simulated regimes: feedback off at $10^{-2}~\mathrm{mbar}$ and feedback on at $10^{-7}~\mathrm{mbar}$. Because both pressure and feedback state differ, the curves illustrate the two operating regimes rather than a one-parameter comparison.

The reduced area of the displacement power spectral density in the cooled regime corresponds to lower motional energy and greater localization around the equilibrium position. This behavior is qualitatively consistent with the spectra reported by Dania \textit{et al.}~\cite{dani2021}.

\begin{figure}[htbp]
    \centering
    \includegraphics[width=0.72\linewidth]{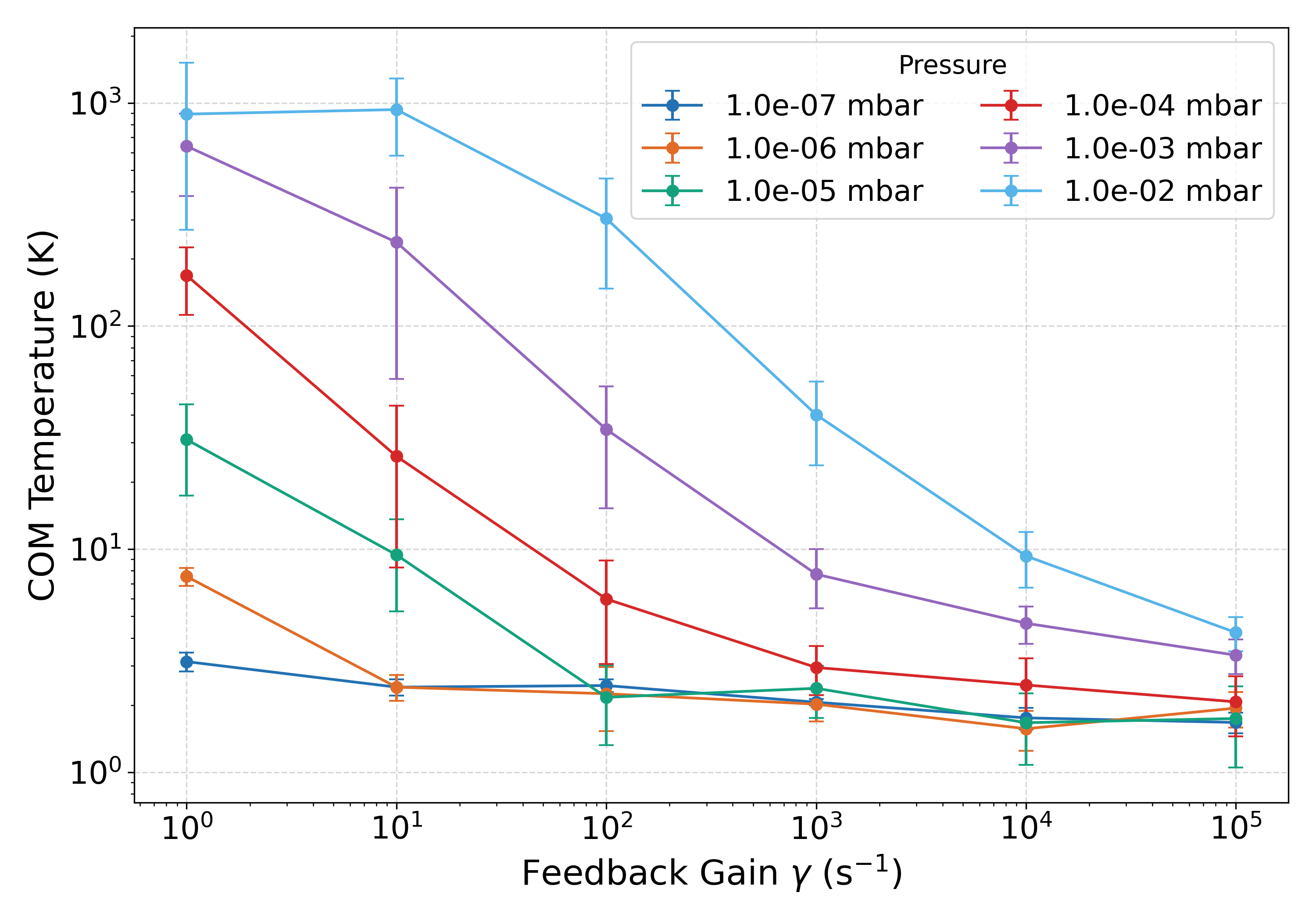}
    \caption{Steady-state COM temperature as a function of feedback damping rate for different background pressures. In the idealized detection model, stronger feedback increases the total mechanical damping and lowers the effective motional temperature.}
    \label{fig:tcom_vs_gamma}
\end{figure}

The steady-state COM temperature was then calculated over a range of pressures and feedback damping rates. As shown in Fig.~\ref{fig:tcom_vs_gamma}, $T_{\mathrm{COM}}$ decreases as $\gamma_{\mathrm{fb}}$ increases, especially once feedback damping exceeds gas damping. Within the simulator's idealized detection model, this trend marks the transition from gas-dominated motion to feedback-controlled dynamics.

The monotonic decrease in Fig.~\ref{fig:tcom_vs_gamma} does not include the measurement-noise limit observed experimentally. In the experiment, increasing the gain initially improves cooling, but beyond an optimum value, measurement noise is amplified by the feedback loop and reinjected into the particle motion, causing the temperature to rise. The simulator can also become unstable at excessively large damping rates, but through a different mechanism: overly strong or discrete feedback corrections can overshoot the particle motion and cease to act as a purely dissipative force. The resulting high-gain heating is therefore qualitatively similar to the experimental upturn but should not be interpreted as a model of measurement-noise backaction.

\begin{figure}[htbp]
    \centering
    \includegraphics[width=0.50\linewidth]{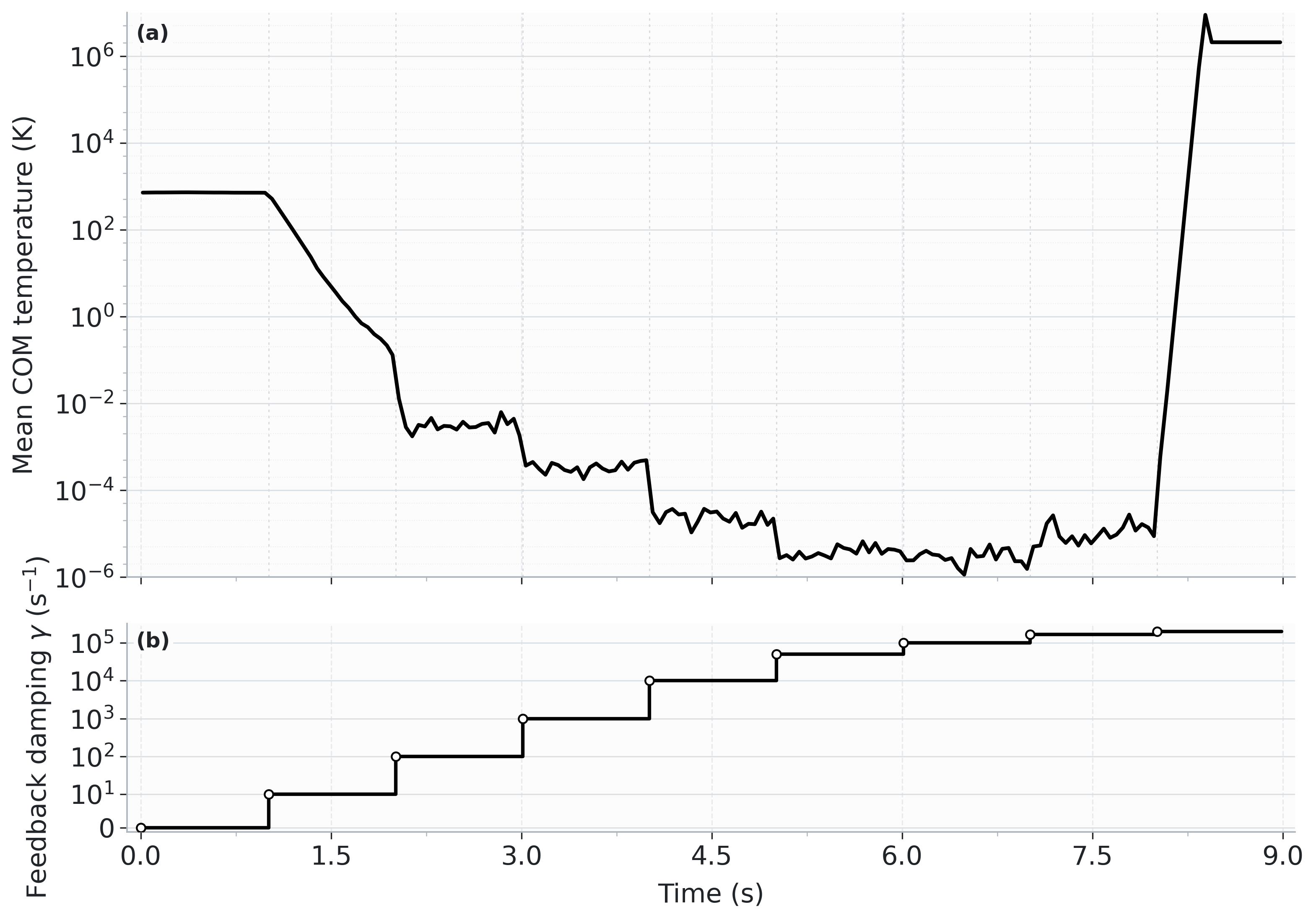}
    \caption{(a) Time-binned mean COM temperature, $T_{\mathrm{COM}}$, at a pressure of $1.0\times10^{-6}~\mathrm{mbar}$, displayed on a logarithmic scale with an averaging window of $0.05~\mathrm{s}$. (b) Applied feedback damping rate, $\gamma_{\mathrm{fb}}$, as a function of time. Vertical dashed lines mark changes in the damping setpoint.}
    \label{fig:tcom_time_gamma}
\end{figure}

Figure~\ref{fig:tcom_time_gamma} illustrates this transition in the time domain. Increasing $\gamma_{\mathrm{fb}}$ initially accelerates cooling and lowers the stationary temperature. At the largest damping rate, however, $T_{\mathrm{COM}}$ rises sharply, indicating that the feedback has entered an energy-injection regime.

\section{Conclusion}

This work presented siNPle, an interactive computational framework for investigating the mechanical, thermal, and spectroscopic dynamics of levitated nanoparticles in Paul traps, optical tweezers, and hybrid trapping configurations. The simulator combines models of trapping forces, residual-gas damping, stochastic motion, feedback cooling, interparticle interactions, internal heating, thermal radiation, optical absorption, fluorescence, and solid-state laser cooling within a unified environment. This integrated approach enables the simultaneous analysis of the center-of-mass and internal dynamics, which are often studied separately despite their potential coupling in realistic experiments.

The simulator provides access to a broad multidimensional parameter space, allowing users to investigate how particle properties, trap parameters, gas pressure, laser power, feedback strength, and spectroscopic characteristics affect the system dynamics. Automated parameter sweeps and data export in CSV and binary formats also facilitate systematic studies and external analysis. Comparisons with representative results from the literature demonstrate that the simulator can reproduce relevant qualitative trends and characteristic dynamical regimes, including pressure-dependent motion, feedback cooling, internal optical heating and cooling, and correlations between the center-of-mass and internal temperatures.

The purpose of siNPle is not to replace detailed electromagnetic, atomistic, or material-specific calculations. Instead, it provides an intermediate level of description between simplified analytical estimates and computationally demanding multiphysics simulations. This makes it particularly useful for identifying dominant physical mechanisms, estimating appropriate experimental parameters, testing hypotheses, and exploring conditions that may be difficult or costly to investigate directly in the laboratory.

Nevertheless, the results must always be interpreted within the assumptions and validity ranges of the implemented models. Some processes are represented phenomenologically, and quantitative predictions may depend strongly on the accuracy of the material properties, optical spectra, gas parameters, and coupling coefficients supplied by the user. Consequently, simulations intended to describe a specific experimental system should be calibrated and compared with experimental measurements whenever possible.

The  siNPle provides a flexible and computationally efficient platform for connecting theoretical models with experimentally accessible quantities. It can therefore support the design, interpretation, and optimization of experiments involving levitated nanoparticles, hybrid Paul--optical traps, feedback control, and solid-state laser cooling.

\end{document}